\documentclass[manuscript]{aastex}

\slugcomment{To be submitted}

\shorttitle{Spiral galaxies gradients}
\shortauthors{Stanghellini et al.}

\begin{document}


\title{Gas-phase oxygen abundances and radial metallicity gradients in the two nearby spiral galaxies NGC~7793 and NGC~4945}


\author{Letizia Stanghellini}
\affil{National Optical Astronomy Observatories, Tucson, AZ 85719}
\email{lstanghellini@noao.edu}

\author{Laura Magrini}
\affil{INAF - Osservatorio Astrofisico di Arcetri, Largo E. Fermi, 5, I-50125 Firenze, Italy}
\email{laura@arcetri.astro.it}

\and

\author{Viviana Casasola\altaffilmark{1}}

\affil{INAF - Osservatorio Astrofisico di Arcetri, Largo E. Fermi, 5, I-50125 Firenze, Italy}
\email{casasola@arcetri.astro.it}
\altaffiltext{1}{INAF - Istituto di Radioastronomia \& Italian ALMA Regional Center, Bologna, Italy}



\begin{abstract}
Gas-phase abundances in H~II regions of two spiral galaxies, NGC~7793 and NGC~4945, have been studied to determine their radial metallicity gradients. We used the strong-line method to derive oxygen abundances from spectra acquired with GMOS-S, the multi-object spectrograph on the 8m- Gemini South telescope.  We found that NGC~7793 has a well-defined gas-phase radial oxygen gradient of -0.321 $\pm$ 0.112 dex R$_{\rm 25}^{-1}$ (or -0.054 $\pm$ 0.019 dex kpc$^{-1}$) in the galactocentric range  0.17$<$R$_{\rm G}$/R$_{\rm 25}$ $<$ 0.82, not dissimilar from gradients calculated with direct abundance methods in galaxies of similar mass and morphology. We also determined a shallow radial oxygen gradient in NGC~4945, -0.253 $\pm$ 0.149 dex R$_{\rm 25}^{-1}$ (or -0.019 $\pm$ 0.011 dex kpc$^{-1}$) for 0.04$<$R$_{\rm G}$/R$_{\rm 25}$ $<$ 0.51, where the larger relative uncertainty derives mostly from the larger inclination of this galaxy.  NGC~7793 and NGC~4945 have been selected for this study because they are similar, in mass and morphology,  to M33 and the Milky Way, respectively. Since at zeroth order we expect the radial metallicity gradients to depend on mass and galaxy type, we compared our galaxies in the framework of radial metallicity models best suited for M33 and the Galaxy. We found a good agreement between M33 and NGC~7793, pointing toward similar evolution for the two galaxies. We notice instead differences between NGC~4945 and the radial metallicity gradient model that best fits the Milky Way. We found that these differences are likely related to the presence of an AGN combined with a bar in the central regions of NGC~4945, and to its interacting environment.
\end{abstract}





\section{Introduction}
The metallicity of galaxies reveals their chemical evolution through cosmic time, and it varies according to star formation history, gas accretion during the early evolutionary stages, and subsequent gas inflows and outflows. The gas-phase metallicity has been studied in the Milky Way and nearby spiral galaxies through abundance analysis of H~II regions (e.g., Vila-Costas \& Edmunds 1992,  S{\'a}nchez et 
al. 2014), disclosing that a radial metallicity gradient across the galactic disk is almost invariably negative (i.~e., higher abundances near the galactic center). The oxygen metallicity gradient is a key constraint to models of chemical evolution. It is in fact essential to pin down this value in the present-day spiral galaxies to assess the gas infall processes. Metal-poor gas infall into the galaxy potential well would lower the central metallicity. Gas accretion through interactions with nearby galaxies would change the gradient as well, compromising the metallicity pattern achieved with stellar evolution and recycling. Even the sophisticated, cosmological-compatible, chemical evolution models of spiral galaxies (e.g., Rahimi et al. 2011; Pilkington et al. 2012; Gibson et al. 2013) cannot give such constraints ab initio. Observational constraints are essential to advance in this field. 

In the recent years, a remarkable body of H~II region abundances has become available to show that resolved spiral galaxies display negative gas-phase radial oxygen gradients. A few galaxies have been studied through direct abundance analysis, which is based on the determination of the electron temperature in the regions from the auroral emission lines. While a discrepancy has been noted between electron temperatures derived from auroral and recombination-lines (e.~g., Pe{\~n}a 2011), this method give arguably the best sets of abundances published to date, resulting in shallow radial oxygen gradients with slopes in a narrow range (-0.02 to -0.09 dex kpc$^{-1}$) for M33 (Magrini et al. 2010), NGC~300 (Bresolin et al. 2009), M81 (Stanghellini et al. 2014), and NGC~628 and NGC~2403 (Berg et al. 2013). The database of galaxies whose H~II region oxygen abundance is known through direct abundance analysis is limited, thus there was purpose to continue this type of studies in a variety of galaxy types. It is also essential to perform abundance analysis in galaxies that are {\it twins} to the ones already well studied. In fact, at first glance, the gradient differences across galaxies can be imputed to the galaxy mass, its morphology, or its situation within a group. As an example, Stanghellini et al. (2014) found that M81 had the steeper gradient slope of all galaxies studied through H~II direct abundance analysis, arguably due to its environment. 

In order to augment the sample of spiral galaxies whose resolved H~II regions can be used to constrain the metallicity gradients, we embarked on a series of multi-object spectrographic studies to detect H~II oxygen abundances in a variety of nearby spiral galaxies. In this paper we present a study of the Sculptor galaxy NGC~7793, and of the Centaurus galaxy NGC~4945. NGC~7793 has apparently sub-solar metallicity (0.6 Z$_{\odot}$, Van Dyk et al. 2012), determined only locally from H II regions.  With a distance of 3.91$\pm$0.41 Mpc (Karachentsev et al. 2003), apparent diameter of 10 arcmin, and intermediate inclination, this galaxy is an ideal target for GMOS multi-object spectroscopy. NGC~7793 is morphologically very similar to M33, it also lacks a defined bulge, and has similar mass and size. Our other target is the less studied NGC~4945,  referred to as Milky Way twin for its similarity with the Galaxy, due to its mass, size, and morphological type.  NGC~4945 is very inclined on the plane of the sky, and has heliocentric distance (3.82 $\pm$ 0.31 Mpc, Lin et al. 2011) similar to that of NGC~7793. The basic physical parameters of both galaxies, and those of their twins, are given in Table~1.

We describe the observations and spectral analysis in $\S$2. In $\S$3 and $\S$4 we present the detailed flux and abundance analysis, both for NGC~7793, and NGC~4945. The study of radial metallicity gradients is given in $\S$5.
The discussion in $\S$6 includes a comparison of the strong- and direct abundances in M33 and M81, and the radial metallicity gradients in NGC~7793 and NGC~4945 compared with chemical evolution models. Our conclusions are given in $\S$7.

\section{Observations and spectral analysis}

In Table~2 we give the GMOS-S observing logs. The NGC~7793 images were acquired in September and October 2012,  while spectroscopic observations of this galaxy have been acquired in July and August 2013. The H~II regions were identified on the images by means of the onband/offband technique, utilizing continuum-subtracted H$\alpha$ images. We built the MOS mask by selecting targets that were bright in the H$\alpha$ continuum-subtracted frames.  H$\alpha$ images of NGC~4945 have been acquired in March 2013, while spectroscopy was obtained in June and July of the same year. We have also acquired [O~III] 5007~\AA~ continuum-subtracted images of both galaxies as part of the original program, aimed at both planetary nebula (PN) and H~II region spectroscopy. [O~III] images can be used to determine PN contamination in the frames. As it turned out, all H~II regions observed were marginally extended in the images, thus they can not be PNe. The [O~III] images will be used in the future to build MOS masks for PN spectroscopy.

In Figures~1 and 2 we show the images of NGC~7793 and NGC~4945. These figures have been obtained form continuum-subtracted H$\alpha$ images. We mark the H~II regions selected for spectral analysis in these images. The spatial distribution of the selected H~II regions, in each galaxy, is adequate for radial metallicity gradient analysis.

The NGC~7793 galaxy has inclination similar to that of NGC~2403, and to both NGC~300 and M33. Its gradient can be reasonably explored through just one GMOS field that includes about half of the galactic face.  In oder to establish the metallicity gradients based on the observed surface of NGC~7793, we have to assume no or little azimuthal variation of gas-phase oxygen, which is consistent with the low metallicity scatter of $\sim$0.05 dex measured in various spiral galaxies to date  (e.g., Bresolin et al. 2009, 2011).

On the other hand, NGC~4945 is nearly edge-on, and its projection results elongated on the plane of the sky, so that we acquired two MOS fields of this galaxy, in order to detect suitable H~II region spectroscopy for metallicity gradients.

Spectroscopy of 27 H~II regions in NGC~7793, and 33 H~II regions in NGC~4945, was obtained with dark sky background (SB50), high cloud coverage (CC70), and poor image quality (IQ70), which yielded to lower quality data than requested for the originally planned science goal, i.e., to determine direct abundances. Our preliminary data analysis confirmed that the data were adequate for strong-line analysis, with S/N$>$3 in all diagnostic lines, unless specifically noted (see Table~4). Even with lowered observing conditions, there was purpose in publishing the present data sets, including line fluxes and their uncertainties, and the strong-line abundances. 

Our observational design is identical to that used in Stanghellini et al. (2014) for the H~II regions of M81 H~II, and details can be found therein. We employed both the R400 and B600 gratings in order to obtain the full spectral domain between 3200 and 9000~\AA, needed for the science goals. We used 1\arcsec~ slitlet widths with 2$\times$2 binning. Data analysis has been performed with the Gemini IRAF routines, as in Stanghellini et al. (2014). Sky subtraction (and, more correctly, galactic background subtraction) was performed by using extended slit margins during spectral extraction, which was possible having designed masks with slitlets lengths larger than the spatial target dimensions. Our target slitlets are 4 to 10 arcsec in length.

\section{H~II region spectroscopy}

In Tables~3  and 4 we list the regions observed in NGC~7793 and NGC~4945 respectively. Preliminary spectral inspection of NGC~7793 showed that S/N was very low in regions 4, 7, and 10, so their fluxes can not be used for abundance determination.  Few more regions (5, 13, and 27) have low quality spectra, thus  only suitable for limited analysis. In Tables~3 and 4 we also give the galactocentric distances of the regions, both in terms of the isophotal radius (Column 4) and in kpc (Column 5).  The uncertainties in the galactocentric distances have been determined with error propagation, including the inclination, position angle, and either the isophotal radius (for R$_{\rm G}$ in terms of R$_{\rm 25}$) or the galaxy distance (for R$_{\rm G}$ in kpc) uncertainties.

In order to confirm that the observed targets are indeed H~II regions we have used the BPT diagram, which is based on Balmer, and on [N~II], [O~III], and [S~II], emission lines (Baldwin et al. 1981; Kniazev et al. 2008).  In Figure 3 we show the [O~III]~$\lambda$5007/H$\beta$ vs. [N~II]$\lambda$6584/H$\alpha$ diagnostic plot for the observed regions in both galaxies. We plot all regions except regions 4, 7, and 10 in NGC~7793. All flux ratios are plotted with their formal errors, propagated from the individual flux uncertainties. Most NGC~7793 targets, indicated with filled circles in the figure, are located below the ${\rm I_{5007}/I_{H\beta} > 0.61 (I_{6584}-0.47)+1.19}$  curve (Kniazev et al. 2008), confirming their H~II nature.  For this galaxy, a few regions are located on or above the curve. Of these, we can confirm that NGC~7793 regions 16 and 25 are likely H~II regions by using the [S~II] line diagnostic from Kniazev et al. (2008), where ${\rm log (I_{6717}/I_{H\alpha}) < 0.63 ~ log (I_{6584}/I_{H\alpha})-0.55}$. On the other hand, regions 14, 17, 23, and 26 in NGC~7793 could be PNe based on the [S~II] diagnostics, and are excluded from further analysis. All NGC~4945 H~II regions studied here are located below the BPT curve, except region 20 of field 2, which we also exclude from gradient analysis. 

Observations of the complete optical spectra were achieved using two gratings per MOS field, specifically, B600 and R400. Since the same emission lines could occasionally be measured in both gratings, the comparison across gratings is a sanity check to determine whether the data reduction has been performed properly.  Also, scaling the red and blue spectra allow us to use diagnostics across the spectra, for example, to determine line extinction. In Figure 4 we show the comparison of the bright lines across gratings, where we plotted the logarithmic fluxes detected in the R400 vs. those in the B600 grating. Both galaxies, and all regions with common lines across gratings, have been included in this plot (we did exclude regions 4, 7, and 10 of NGC~7793, see above).  We used the brightest common line in each region (typically H$\alpha$, [O III] $\lambda$5007, or [N II] $\lambda$6584 \AA) to make the grating comparison. We found that the line fluxes across gratings are generally well correlated in all fields studied. By excluding targets of uncertain nature, and those with low S/N spectra, we obtain linear correlation index R$_{\rm xy}$=0.98 (in the logs) in NGC~7793, and 0.995 and 0.991 respectively for fields 1 and 2 in NGC~4945. We used the flux ratio of the brightest lines detected in both gratings to scale the red and blue fluxes. For regions with no lines in common across gratings, we used the average flux ratios from the same field to scale lines across gratings. The average flux ratio for NGC~7793, and for field 1 and field 2 of NGC~4945, are respectively $<{\rm F_{B600}/F_{R400}}>=1.388\pm0.556$, 1.25 $\pm$ 0.16, and 1.09 $\pm$ 0.22.

Line fluxes were measured on the reduced and scaled spectra with the IRAF routine {\it splot}. Flux uncertainties were carried over from these measurements. We determined extinction correction, following Osterbrock \& Ferland 's (2006) prescription for T$_{\rm e}$=10,000 K, for most regions. In Tables~5 and 6 (available electronically) we list the measured emission lines scaled to H$\beta$=100, respectively for H~II regions in NGC~7793 and NGC~4945. We give name, logarithmic extinction correction, and H$\beta$ flux first, then  we list all emission lines by giving the wavelength (Column 1), line identification (Column 2), and observed flux (Column 3) and uncertainty (Column 4), and finally the de-reddened flux (Column 5). In a few cases the extinction turns out to be negative, and we set it to zero. 

We requested to observe all MOS fields with the most suitable parallactic angle. Instead, most of the spectra were acquired with parallactic angle $\sim 90$ deg, thus the lines near the blue end of the B600 spectra are not usable. In particular, the blue [O~II] 3727-3729~\AA~ lines, where present, have low S/N. We deem those lines not suitable for abundance analysis. In this paper we publish only the strong line fluxes relative to the abundance analysis presented here.

\section{Strong-line abundances}

Gas-phase oxygen abundances are best derived directly from line fluxes and electron densities and temperatures of the gas.  Our data set suffers from low S/N auroral lines, making the direct analysis unfeasible, allowing instead the determination of oxygen abundances via strong-line diagnostics. The fundamental strong lines in our spectra that can be used for abundance diagnostics are H$\beta$, [O~III] $\lambda\lambda$4959, 5007, H$\alpha$, [N~II] $\lambda\lambda$6548, 6584 , and [S~II] $\lambda\lambda$6717, 6731~\AA.  The various diagnostic methods use all, or a partial suite of, these lines; maximizing the number of diagnostic lines helps to break degeneracy between abundances and ionization parameters. 

We took the opportunity offered by the many emission lines observed and measured to obtain abundances from several strong-line methods. The availability of the [N~II]  6584~\AA~ and [S~II] $\lambda\lambda$6717, 6731~\AA~ lines, together with the strong [O~III] and the optical Balmer lines, allow the determination of the ionizing field in the absence of [O~II] observations. In the cases where [O~III], [S~II], [N~II], H$\alpha$, and H$\beta$ were all simultaneously available, we determined the oxygen abundances based on the O3S2N2 (also known as {\it NS}) method by Pilyugin \& Mattsson (2011). We followed their prescription at different metallicities (i.e., Equations 8 in Pilyugin \& Mattsson 2011) and give the resulting abundances in Column (2) of Tables~7 and 8 (respectively for NGC~7793 and NGC~4945).  

The O3N2\ index can be used even when the [S~II] lines are either too faint, or not available in the spectra. This method is based on the empirical correlation between the O3N2 flux ratio, defined as 
O3N2=log ([O~III] $\lambda5007$/H$\beta$ / [N~II]$\lambda6584$/H$\alpha$),  and the oxygen abundance. Pettini \& Pagel (2004) and Charlot \& Longhetti (2001, case E) have calibrated this index. More recently, Marino et al. (2013) have used a large number of calibrators with known T$_{\rm e}$ to determine a new correlation between the O3N2 index and the oxygen abundances of hundreds of extragalactic  H~II regions, and were able to lower calibration uncertainties with respect to those quoted by previous Authors. We used Eq. (2) in Marino et al. (2013) to determine the O3N2 abundances presented here. The O3N2 abundances, calculated only if O3N2 is within the domain of the correlation (Marino et al. 2013), are given in Column (3) of Tables~7 and 8. 

If only [N~II] and H$\alpha$ are available, a rough estimate of the oxygen abundances can be obtained by fitting the direct oxygen abundances vs. the N2 (log[N~II]$\lambda6584$/H$\alpha$) index (see e.g,, Pettini \& Pagel 2004; Denicol{\'o} et al. 2002). We calculated the N2 abundances following the new N2 calibration by Marino et al. (2013, Eq. 4). We give the N2 abundances in Column (4) of Tables~7 and 8 for all cases where the index N2 is within the domain of the calibration. It is worth noting that strong-line abundances from both the O3N2 and N2 methods do not suffer from uncertainties associated with cross-grating calibrations, and that the extinction corrections have a small impact on these calibrations, since they utilize line ratios that are close in wavelength. 

The [O~II] emission lines at $\lambda$3727~\AA~ are available for several regions, but their fluxes should be considered lower limits, and thus not useful to determine abundances with the O2 prescription by Kewley \& Dopita (2002). 

In Figure~5 we compare abundances based on the O3N2 (filled symbols) and N2 (open symbols) indexes vs. those obtained with the O3S2N2 calibration, for regions in both galaxies (circles for NGC~7793, squares for NGC~4945). Here we exclude all regions that are flagged for low S/N ($<$2$\sigma$) or uncertain identification.  By calculating the residuals between the O3S2N2 abundances and those obtained with the O3N2 and N2 indexes, we find 
$< {\rm log}(O/H)_{\rm O3S2N2} - {\rm log}(O/H)_{\rm O2N2}>$= 0.07$\pm$0.072, and  $<{\rm log}(O/H)_{\rm O3S2N2} - {\rm log}(O/H)_{\rm N2} >$=0.11$\pm$ 0.08, both based on 30 targets. The linear correlation coefficients (in the logs) are R$_{\rm xy}$=0.81 and 0.71 respectively for the O3N2 and N2 comparisons. 

A further quantitative assessment of the different abundance calibrations is given in the discussion. While the O3S2N2 method is to be preferred, being based on three flux ratios that are sensitive to both ionization and abundance, it is also the one that most suffer from the correction across gratings, since all red line fluxes are divided by H$\beta$ to produce the indexes. The O3N2 method also suffers from the calibration across gratings. 

The abundance uncertainties in Tables~7 and 8 are calculated by error propagation, and include both the calibration uncertainties from the indexes, both random and systematic, and the flux uncertainties from our measurements.

\section{Radial oxygen gradients, and characteristic abundances}

We determine the radial metallicity gradients for both galaxies by fitting the (logarithmic) oxygen abundances vs. the galactocentric distances with the {\it fitexy} routine in Numerical Recipes (Press et al. 1988). We calculate the gradients both in dex R$_{\rm 25}^{-1}$, and in dex kpc$^{-1}$. We included in these fits both the galactocentric distance and abundance uncertainties; the abundance uncertainties have been derived by error propagation as described in the previous section; uncertainties in the galactocentric distance are estimated by propagating the errors in inclination, position angle, and R$_{\rm 25}$ (or the galaxy distance, if the gradient is given in dex kpc$^{-1}$). 

In Table~9 we give the calculated gradients and average abundances for both galaxies. In Column (1) we give the abundance calibration index; Column (2) gives the number of data points available; Column (3) gives the galactocentric range of the data points, in terms of R$_{\rm 25}$; Columns (4) and (5) give the gradient slope, respectively in dex R$_{\rm 25}^{-1}$ and in dex kpc$^{-1}$, with their uncertainties; column 6 gives the intercept of the fit, and its uncertainty; and finally, Column (7) gives the average abundance in the radial range of Column (3).

It is evident from Table~9 that the average abundances are higher for the O3S2N2 index than for the other ones, in both galaxies, by a factor of $\sim$1.1 to 1.4. Note that all averages have been calculated for the same radial range within each galaxy.

In Figure~6 we show the radial metallicity gradients derived for NGC~7793 H~II regions with the three indexes. We show the results from the different calibrations in three different panels, for clarity (see figure caption), and determine radial oxygen gradients from homogeneous data sets (i.e., from the same calibration). In the gradient analysis we excluded all flagged regions of Table~3, thus we also exclude probable (but not confirmed) H~II regions (which are instead included in the abundance comparison of Figure~5). 

As a comparison, and to show the improvement of our data set over the existing literature, we have considered the line fluxes of H~II regions in NGC~7793 from Bibby \& Crowther (2010) and Webster \& Smith (1983, as corrected by Moustakas et al. 2010) whose oxygen abundances we recalculate following the N2 method (Marino et al. 2013), and found a gradient slope $\sim$-0.270 dex R$_{\rm 25}^{-1}$. The gradient described by our own abundances have considerable less scatter (The O3S2N2 calibration gives a linear correlation R$_{\rm xy}\sim$-0.8) than that from the existing literature data
(R$_{\rm xy}$=-0.6). NGC~7793 was also analyzed within the data sets included in the comprehensive work by Pilyugin et al. (2014). They found a radial oxygen gradient -0.305 $\pm$ 0.048 dex R$_{\rm 25}^{-1}$, which is consistent with our result, within the errors.
Figure~7 is similar to Figure~6, but for NGC~4945 H~II regions. All regions not flagged in Table~4 are included in the gradients. We plot radial oxygen gradients from the different abundance calibrations, where the galactocentric distances are form Table~4 and the abundances from Table~8.

\section{Discussion}

\subsection{Testing the reliability of strong-line abundance calibrations: Comparing strong and direct abundances of M81 and M33 H~II regions}

We have shown that strong-line oxygen abundances of H~II regions in NGC~7793 and NGC~4945 yield to shallow radial metallicity gradients, with negative slopes. An additional check of the different calibration can be obtained by analyzing H~II regions in M33 and M81, where direct abundances are available for large samples of H~II regions. These two galaxies have recently been analyzed in terms of their direct gas phase oxygen abundances (Stanghellini et al. 2014, Magrini et al. 2010), and  have rather different metallicities, thus are excellent probes of the methods.

We thus selected line fluxes of M81 and M33 H~II regions from Stanghellini et al. (2014) and Magrini et al. (2009a), and calculated strong-line abundances with the O3S2N2, O3N2, and N2 indexes. We then compared strong-line and direct abundances.  We choose targets whose fluxes in the T$_{\rm e}$ diagnostic lines had strengths $> 2\sigma$. By correlating O3S2N2-calibrated and direct abundances, based on 28 common targets, we found that the average of the residuals is $< |{\rm log}(O/H)_{\rm direct}-{\rm log}(O/H)_{\rm O3S2N2}|>$=0.18 $\pm$ 0.16. 
By comparing O3N2 calibration with direct abundances we obtained $< |{\rm log}(O/H)_{\rm direct}-{\rm log}(O/H)_{\rm O3N2}|>0.20\pm0.14$, based on 27 targets. For abundances calibrated with N2 we could compare 32 targets, and found $< |{\rm log}(O/H)_{\rm direct}-{\rm log}(O/H)_{\rm N2}|>=0.20\pm0.13$. The O3S2N2 calibration gives the lowest residuals with respect to the direct abundances, and should be preferred when available. This conclusion does not depend on the selected 2$\sigma$ limit for the diagnostic line fluxes. In fact, by choosing line strengths $> 3\sigma$  we would obtain that the O3S2N2 calibration has the smaller residuals, on average.

In Figure 8 we plot the direct (open symbols) and O3S2N2-derived (filled symbols) abundances against the O3N2 index, as derived from observations of M33 (circles) and M81 (squares) H~II regions. Direct abundances plotted here are based on temperature diagnostic lines with strengths $> 2\sigma$. The solid line represents the O3N2 calibration, and the broken lines its range of uncertainty, as given in the original paper( $\pm$ 0.36 dex, where, according to Marino et al., 95$\%$ of the data points are to be found). It is worth noting that the M33 and M81 abundances were not included in Marino et al's calibrations. A similar plot, but for the N2 calibrator, is shown in Figure 9. From the figures, we noted that the O3N2-method predicts abundances that are lower than both direct and O3S2N2-derived abundances, while the N2-derived abundances are more scattered but seem to trace both the direct and the O3S2N2 abundances.

\subsection{The radial metallicity gradients of NGC~7793 and NGC~4945 as compared to those of their twins}

As described in the introduction, NGC~7793 and NGC~4945 are two important test galaxies because of their strong similarities with the well-studied M33 and the MW.
As seen earlier, Table~1 gives the main properties of the each galaxy and of its closer twin. From the characteristics shown in Table~1 we see that M33 and NGC~7793 are both flocculent late type spiral galaxies, with similar radii and baryonic masses. The MW is slightly brighter and more massive than NGC~4945, but they belong to the same morphological class. Our aim is to compare their strong-line oxygen gradients to see if their intrinsic characteristics --mass and morphology-- are enough to guarantee that their present-time gradients, as traced by H~II regions, are comparable with those of their twins.  

To do that, we contrast the observed metallicity gradients of NGC~7793 and NGC~4945 with the results of two chemical evolution models designed to reproduce the main observational features of M33 and of the MW, respectively. The two models are presented in Magrini et al. (2007, 2010) for M33 and in Magrini et al. (2009b) for the MW. They are part of the family of the multiphase models in which the formation and destruction of diffuse gas, clouds, and stars are followed  by means of the simple parameterizations of physical processes (e.g., Ferrini et al. 1992). The details and equations of the models, together with the specific parameters adopted for the two galaxies, can be find in the original papers. For both galaxies, their model oxygen gradients are constrained with the abundances of H~II regions obtained by the direct measurement of their electron temperature, i.e., by direct oxygen abundances. 

The comparisons of our data with the models are shown in Figure~10, where we plot the gradients from our O3S2N2 abundance analysis to the models of M33 and the Milky Way.  In the top panel we show the observed oxygen abundances from the O3S2N2 calibration against the M33 model gradient (B1 model,  t=0, time scale of the infall=0.003$\times10^{-7}$ [yr${-1}$], coefficient of star formation by cloud collisions= 0.06$\times10^{-7}$ [yr$^{-1}$], coefficient of cloud disruption by cloud collisions= 0.2$\times10^{-7}$ [yr$^{-1}$],
coefficient of cloud formation by condensation of diffuse gas=0.01$\times10^{-7}$ [yr$^{-1}$], Magrini et al. 2007). The model curve of the oxygen gradient of M33 is built to reproduce the direct oxygen abundances of H~II regions. The good agreement between the  O3S2N2-gradient of NGC~7793 with the M33 gradient from the basic model points towards a very similar evolution and chemical enrichment history for the two galaxies. As shown in Table~1, the two galaxies are indeed very similar in many aspects, as well as in their relative isolation within the groups of which they are part (the Local Group for M33, and the Sculptor Group from NGC~7793; Radburn-Smith et al. 2011). The average of the residuals between the two radial metallicity gradients plotted in Fig.~10 is $<$0.06 dex.

The situation shown in the bottom panel of Figure~10, representing NGC~4945, is different. To the O3S2N2-derived  abundances we overplot the basic MW model by Magrini et al. (2009b, see their Fig. A3, present-time model, plotted here with a black solid line. In this case star formation is driven by cloud-cloud collisions, with time scale of the infall= 0.012$\times10^{-7}$ [yr$^{-1}$], coefficient of star formation by cloud collisions= 0.2$\times10^{-7}$ [yr$^{-1}$], coefficient of cloud disruption by cloud collisions= 1.0$\times10^{-7}$ [yr$^{-1}$], coefficient of cloud formation by condensation of diffuse gas=0.09$\times10^{-7}$ [yr$^{-1}$]). The observed O3S2N2 oxygen abundances in NGC~4945, while producing  a shallow gradient, also give lower metallicity that what observed in the MW.  In particular the few regions toward the galactic center do not trace the MW model. We recall that the MW model gradient is constrained with the H~II region oxygen abundances derived by Deharveng et al.~(2000) with the direct method.   

We have thus considered the astrophysical differences between the MW and NGC~4945.  Analyzing the main features of the two galaxies, we found that their diversities are related to the nuclear activity and to the environment.  The central region of NGC 4945 contains an active galactic nucleus (AGN) revealed most unambiguously by its strong and variable hard X-ray emission (Iwasawa et al. 1993; Done et al. 1996; Madejski et al. 2000). The AGN is classified as Seyfert 2 (Braatz et al. 1997; Madejski et al. 2000; Schurch et al. 2002), and is a strong radio continuum source (e.g., Whiteoak \& Wilson 1990). Surrounding the AGN, there is an inclined circumnuclear starburst ring with a radius of $\sim$2.5$^{\prime\prime}$ ($\sim$46 pc), seen most clearly in Pa$\alpha$ (Marconi et al. 2000).  The central region of NGC~4945 is among the strongest and most prolific extragalactic sources of molecular lines (e.g., Henkel et al. 1994; Curran et al. 2001; Wang et al. 2004; Chou et al. 2007).  For these reasons, NGC 4945 is a particularly attractive candidate for studying the nature of molecular gas at the center of an active galaxy, and the role this gas plays in fueling. On the other hand, the MW does not presently appear to be hosting an AGN, given the lack of accretion activity in spite of the presence of a super massive central black hole ($\sim10^{6} M_{\odot}$). It is worth noting that the Fermi Gamma-ray Space Telescope revealed two large gamma-ray bubbles in the Galaxy, which extend about 50 degrees ($\sim$10 kpc) above and below the Galactic center and are symmetric about the Galactic plane (Guo  \& Mathews 2012).  A recent ($\sim$ 1--3 Myr ago) Galactic AGN jet activity of a duration of $\sim$0.1--0.5 Myr could have created these bubbles, as follows from the comparison between these Fermi data and axisymmetric hydrodynamic simulations. Nonetheless, any AGN presence seem to be confined in the Galaxy past.

In addition, NGC~4945 belongs to the Centaurus A Group of galaxies, one of the closest groups of galaxies outside the Local Group whose composition is, however, much different from that of the Local Group, being formed by a heterogeneous assembly of early- and late-type galaxies, all of them characterized by a period of enhanced star formation (C{\^o}t{\'e} et al. 2009). Both the AGN nature and the interaction within a galaxy group might have some importance in shaping the radial metallicity gradient of NGC 4945 though the induction of gas flows along the galactic disk. 
Among the main mechanisms that produce gas flows, there are indeed: \textit{i)} the large quantity of gas around the AGN combined with the presence of one or more bars that produce a falling of gas into the central regions (AGN fueling, e.g., Athanassoula 1992; Friedli et al. 1994; Casasola et al. 2008, 2011; Perez et al. 2011; Combes et al. 2013) and/or \textit{ii)} the interaction of galaxies that induces gas flows from the outer parts to the centre of each component (Toomre \& Toomre 1972; Casasola et al. 2004; Dalcanton 2007).

From the observational point of view, radial inflows, that fuel active galactic nuclei together with the central starburst, might lead to a flattening of the metallicity gradient by gas mixing (e.g., Onodera et al. 2004).  
In this framework, the comparison of NGC~4945 with our Galaxy is, thus, of great importance because we have the possibility to compare a MW twin --in terms of mass and morphology-- but having a central AGN and inhabiting a different environment. Although NGC~4945 resides in a galaxy group, it has not any close massive companion, thus we suspect that the origin of the differences between the gradients of the two galaxies is mainly  related to the presence of an AGN and a bar that induce a continuous infall of material towards the central regions. The presence of a central bar in NGC~4945 has been known for some time (see e.g., Peterson 1980; Lin et al. 2011). From spectroscopic investigation of the ionized gas, Peterson (1980) found strong indication of in falling gas on the southwest side of the galaxy, inferred from departure from purely circular motions. 

Starting from the chemical evolution model of the MW as described in Magrini et al. (2009b), we sketch the effect of an AGN combined with a central bar, as observed in NGC~4945, in two different ways: {\em i)} assuming AGN feedback that reduces 
the star formation efficiency in the inner part of the NGC~4945 disk (e.g., Weinmann et al. 2006),  and  {\em ii)} considering a higher and more continuous infall of metal-poor material, necessary to fuel the AGN activity. 
We thus modified some of the MW model parameters accordingly. In particular,  we considered a model with half the star formation efficiency than that of the MW, and one with an infall rate more diluted in time, corresponding to a longer infall time and a higher infall rate at present-time. The results of these two models in terms of the radial oxygen gradients are shown  in Figure~10, bottom panel, where we plot the low-star forming efficiency model with a light blue solid line, and the larger infall model with a dashed line. The lower star formation efficiency has the effect to produce a smaller disk,  where the metallicity is rapidly declining for regions at galactocentric distances R$_{\rm G}$/R$_{\rm 25}>$0.3, which is not observed in NGC~4945. 
On the other hand, the longer time-scale for the infall produce a shallow metallicity gradient, and a metallicity offset which is consistent with our NGC~4945 observations.  
The model gradient is still more peaked in the central 0.1 R$_{\rm G}$/R$_{\rm 25}$ than the observed one. 

However, the dynamical effect of the
bar, not considered in the chemical evolution model shown here, may have a mixing effect that further flattens the inner gradient, and reduces the overall metallicity because of the
dilution by primordial gas inflowing from the outer disk (see, e.g., Portinari \& Chiosi 2000) and/or  by  AGN-driven super-winds driving gas out of the central regions (Narayanan et al. 2008).
Since our model is a simple semi-analytical chemical evolution model that does not includes dynamical effects, 
we can only speculate that the differences between NGC~4945 and the MW gradients are induced by a presence of gas flows due to both AGN-driven winds and to the bar.

\section{Conclusions}

The strong-line abundance calibrations available to date can give a handle to calculate radial oxygen gradients in spiral galaxies. While more uncertain than the direct methods, if chosen appropriately, they can give constraints to evolutionary models. Strong-line and direct abundances cannot be compared directly in the studied galaxies, since our observations were not deep enough to obtain electron temperatures of the H~II targets. Strong line fluxes allow us to determine abundances by using three different calibrations: the O3S2N2 calibration, and the O3N2 and N2 calibrations. 

To probe the quality of strong-line abundances, we used the available flux data of M33 and M81 H~II regions, and their direct abundances. Comparison between direct and strong-line abundances for these samples indicates that the better calibration is the O3S2N2 index, the one with the larger number of constraints. The O3S2N2 index gives average abundance of (2.55 $\pm$ 0.46)$\times$10$^{-4}$ for 13 NGC~7793 regions in the 0.17$<$R$_{\rm G}$/R$_{\rm 25}$ $<$0.82 radial range, and (3.89 $\pm$ 0.37)$\times$10$^{-4}$ for 17 NGC~4945 regions in the 0.044 $<$ R$_{\rm G}$/R$_{\rm 25}$ $<$ 0.51 radial range.

We obtained a good fit to the abundances and galactocentric distances of NGC~7793, yielding to a radial metallicity gradient with slope of -0.321 $\pm$ 0.112 dex R$_{\rm 25}^{-1}$.  This gradient compares well with that of its galactic twin, M33, which is similar in mass and morphology. The two gradients differ in ordinate less  0.05 dex in the whole radial range for which the NGC~7793 gradient is defined, and with average residuals between the two gradients of 0.029$\pm$0.016 dex, pointing towards a similar chemical evolution for those two galaxies. 

The gradients of NGC~4945 and of the Galaxy show, instead,  a substantial difference in the absolute scale of their metallicity, with an offset of about $\sim$0.15~dex at the galactic center, which can not be accounted for entirely by calibration uncertainties. We have analyzed the differences between NGC~4945 and the MW, and considering the effects of the presence of an AGN combined with a central bar in NGC~4945, which might have primary importance in shaping the radial gradient. Using the multiphase chemical evolution models developed by Magrini et al. (2009b) for the MW, we have varied the infall rate according to the AGN and bar observations, finding that we are able to reproduce a milder gradient with a lower metallicity for NGC~4945, assuming a more continuous inflow of material than for the MW.  We conclude that, while mass and morphology are among the main driver of the shape of the radial metallicity gradient, other aspects, as the presence of a bar, or central activity, or of a close interacting companion, can lead to substantial changes in the shape and/or absolute scale of the metallicity gradients. 
	
The study of twin pairs of galaxies is at its earlier stages, and we plan to continue to use this powerful tool to assess chemical evolution of nearby spiral galaxies via gas-phase metallicity analysis. The oxygen abundance analysis of NGC~4945 would be more effective if the S/N of the spectra would allow direct abundances.  We plan more observations along these lines in the future.

\acknowledgments
We thank Marcel Bergmann for his help in implementing the Phase 2 of this program, and Katia Cunha and Arjun Dey for scientific discussions. We also thank an anonymous Referee for very useful suggestions to an earlier version of this paper.
Based on observations obtained at the Gemini Observatory, which is operated by the 
Association of Universities for Research in Astronomy, Inc., under a cooperative agreement 
with the NSF on behalf of the Gemini partnership: the National Science Foundation 
(United States), the National Research Council (Canada), CONICYT (Chile), the Australian 
Research Council (Australia), Minist\'{e}rio da Ci\^{e}ncia, Tecnologia e Inova\c{c}\~{a}o 
(Brazil) and Ministerio de Ciencia, Tecnolog\'{i}a e Innovaci\'{o}n Productiva (Argentina).
This research has made use of the NASA/IPAC Extragalactic Database (NED) which is operated by the Jet Propulsion Laboratory, California Institute of Technology, under contract with the National Aeronautics and Space Administration. We acknowledge the usage of the HyperLeda database (http://leda.univ-lyon1.fr).



{\it Facilities:} \facility{Gemini}

\clearpage



\begin{figure}
\centering
\title{H$\alpha$ image of NGC~7793}
\includegraphics[width=\hsize]{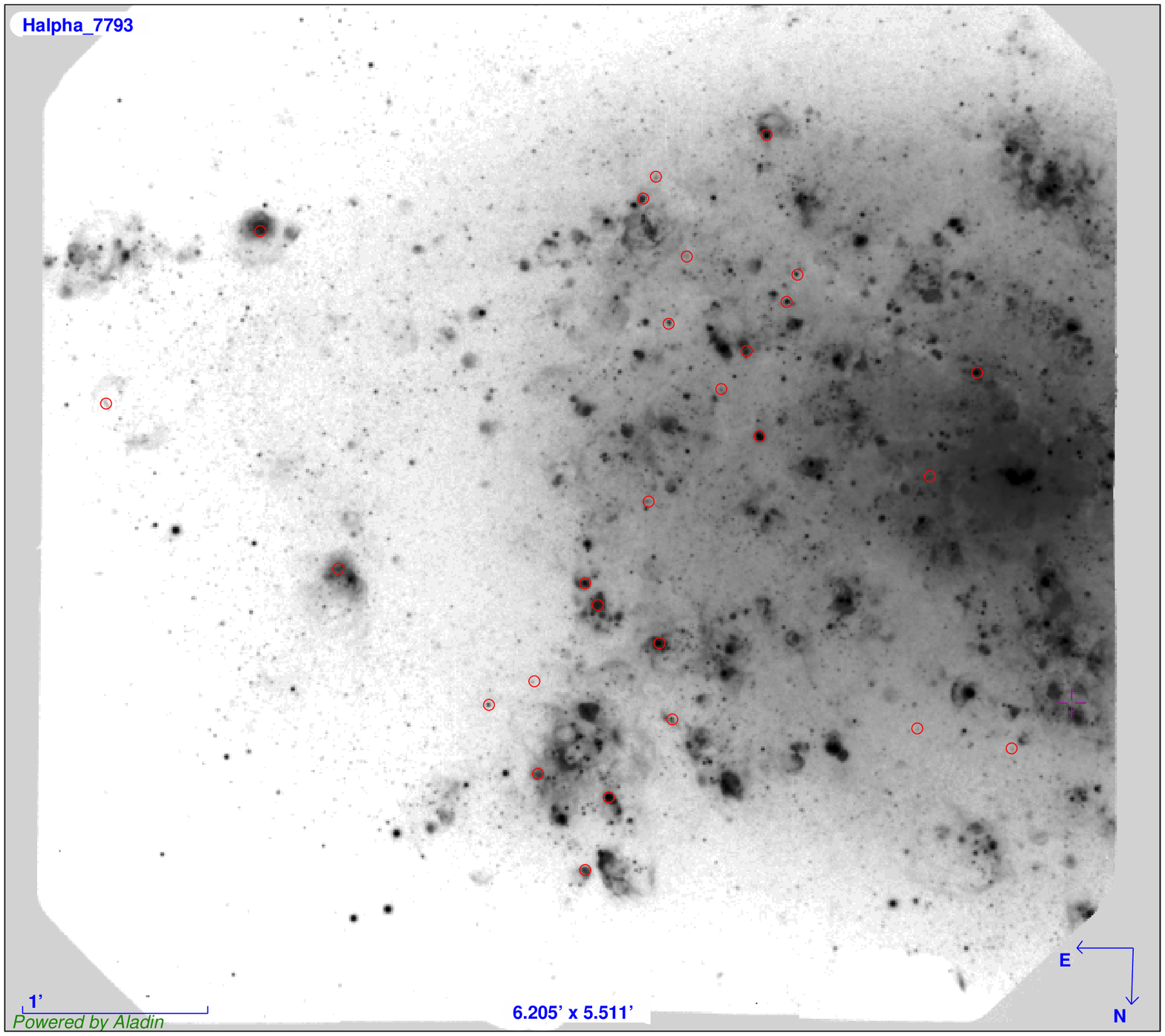}
\caption{H$\alpha$ image of the NGC~7793 field. The galaxy center is to the center right. The (red) circles represent the studied H~II regions.}
\end{figure}

\begin{figure}
\centering
\title{H$\alpha$ images of the two observed fields of NGC~4945}
\includegraphics[width=0.5 \hsize]{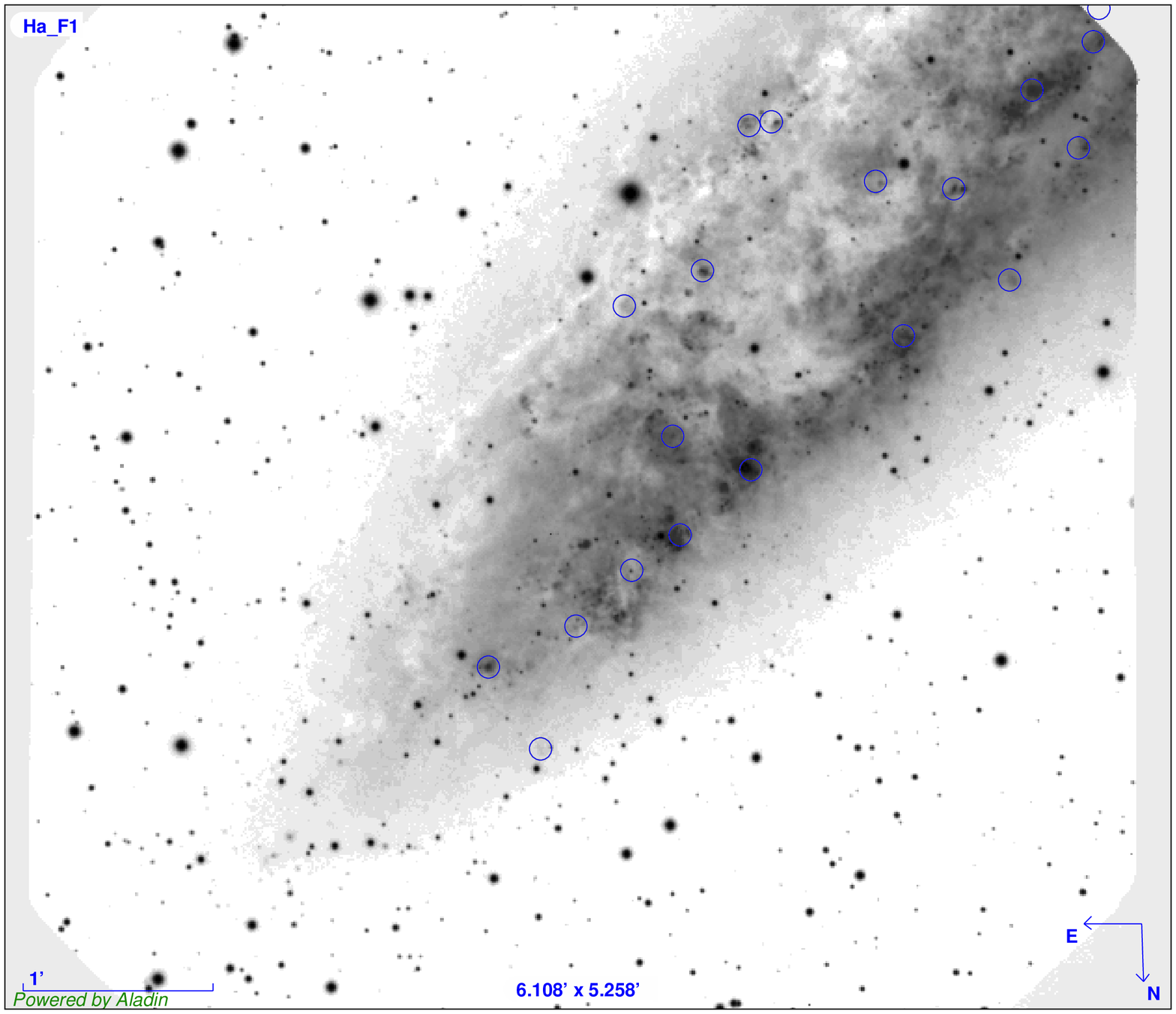}
\includegraphics[width=0.5 \hsize]{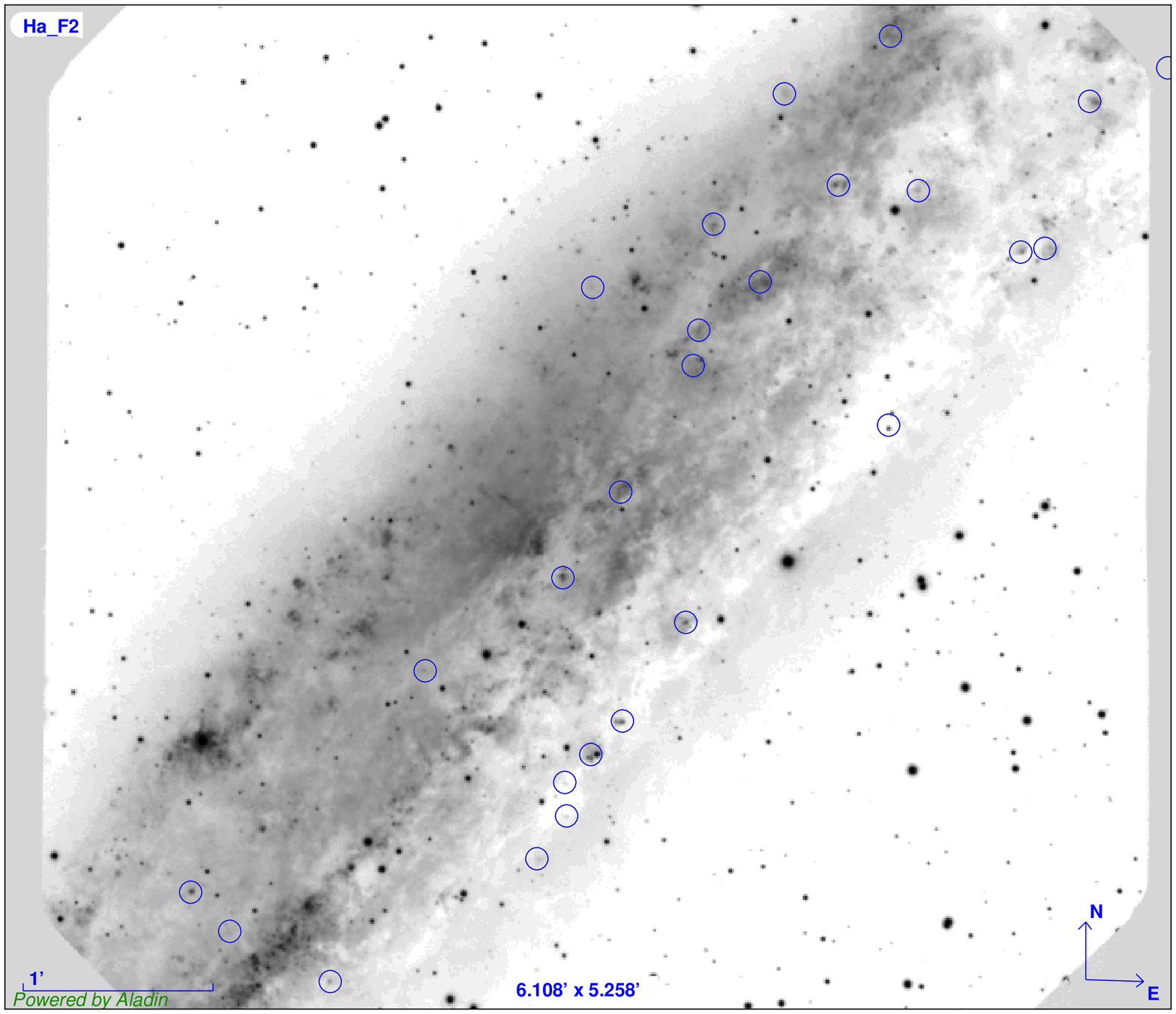}
\caption{H$\alpha$ images of the two fields of  NGC~4945.  The (blue) circles represent the studied H~II regions.}
\end{figure}

\begin{figure}
\centering
\title{BPT diagram of NGC~7793 and NGC~4945 H~II regions}
\includegraphics[width=\hsize]{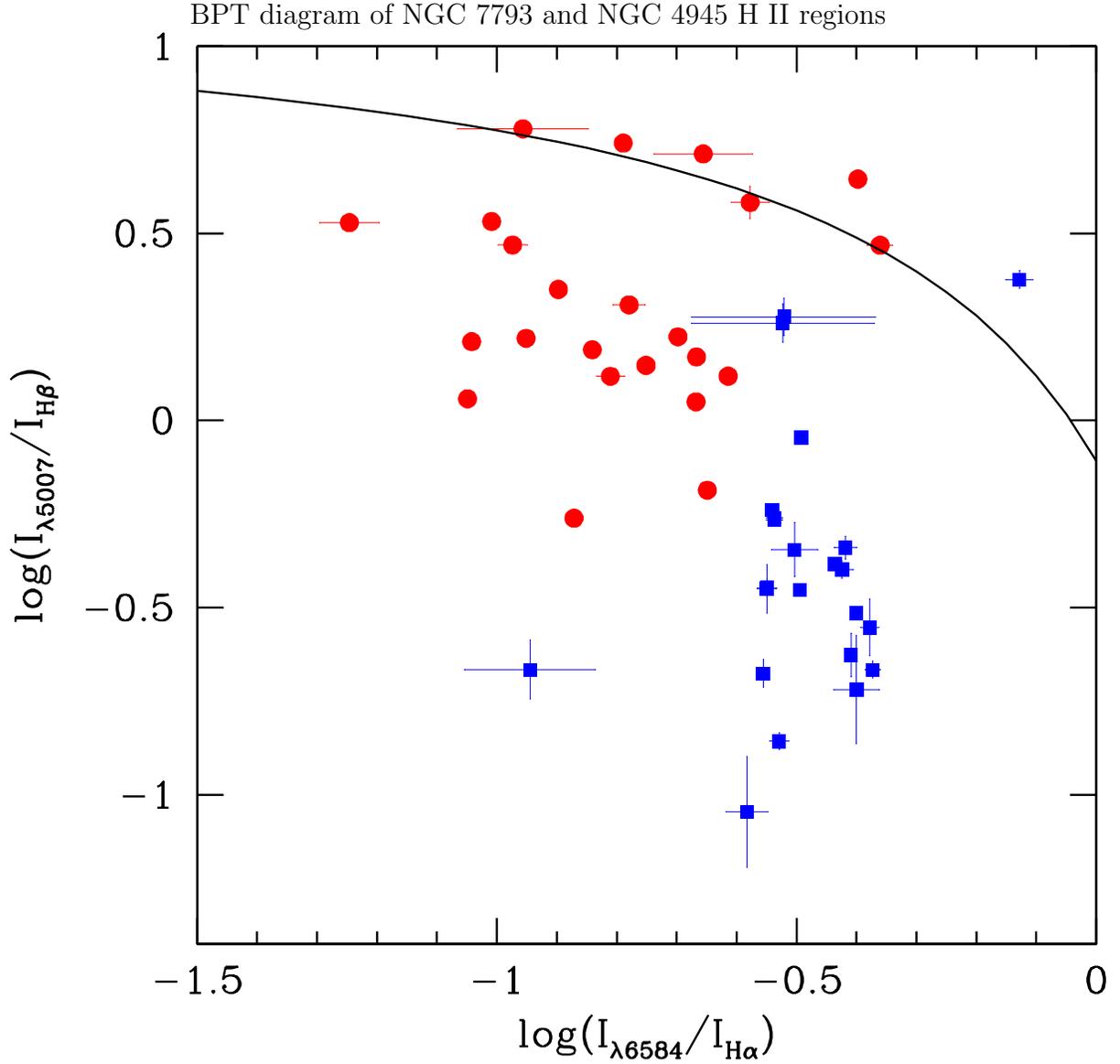}
\caption{Observed regions in NGC~7793 (red filled circles) and NGC~4945 (blue filled squares), in the log(I$_{\rm 6584}/{\rm I_{H\alpha}})$ vs. log(I$_{\rm 5007}/{\rm I_{H\beta}})$ plane. The solid line represents the limiting relation for H~II regions (H~II regions are located below the curve). }
\end{figure}

\clearpage

\begin{figure}
\centering
\title{Bright emission line grating comparison}
\includegraphics[width=\hsize]{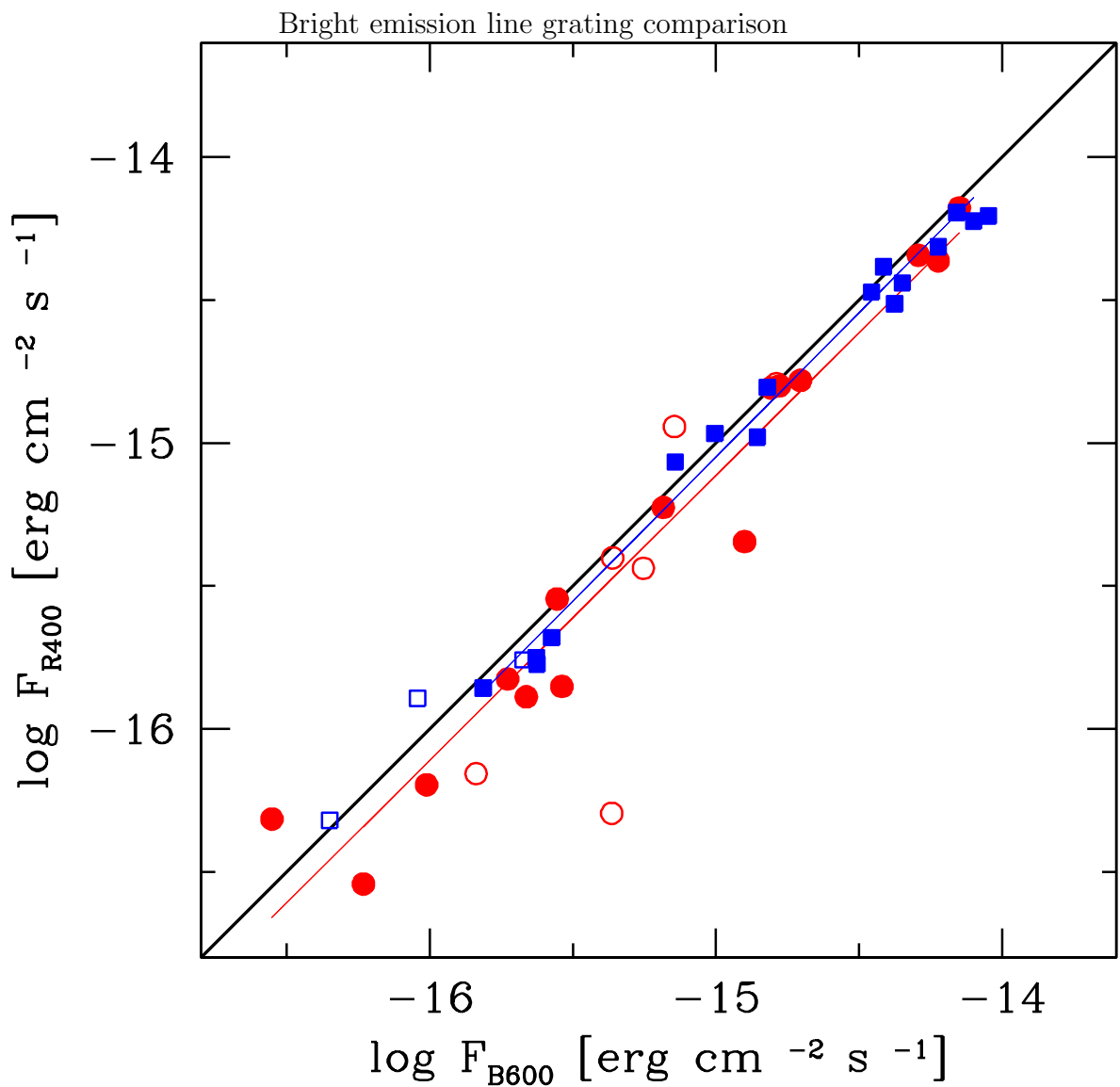}
\caption{Comparison of bright emission lines detected in R400 and B600 for the same targets. NGC~7793 (red filled circles: high S/N spectra of confirmed H~II regions; open symbols: other regions) and NGC~4945 (blue filled squares: high S/N spectra of confirmed H~II regions; open symbols: other regions). The red and blue lines are the fits (high S/N only) for NGC~7793 and NGC~4945, field 2 (the fit for field 1 overlaps that of NGC~7793). The thick solid line is the 1:1 relation.}
\end{figure}

\begin{figure}
\centering
\title{Comparison of O3N2, N2 abundances vs. O3S2N2}
\includegraphics[width=\hsize]{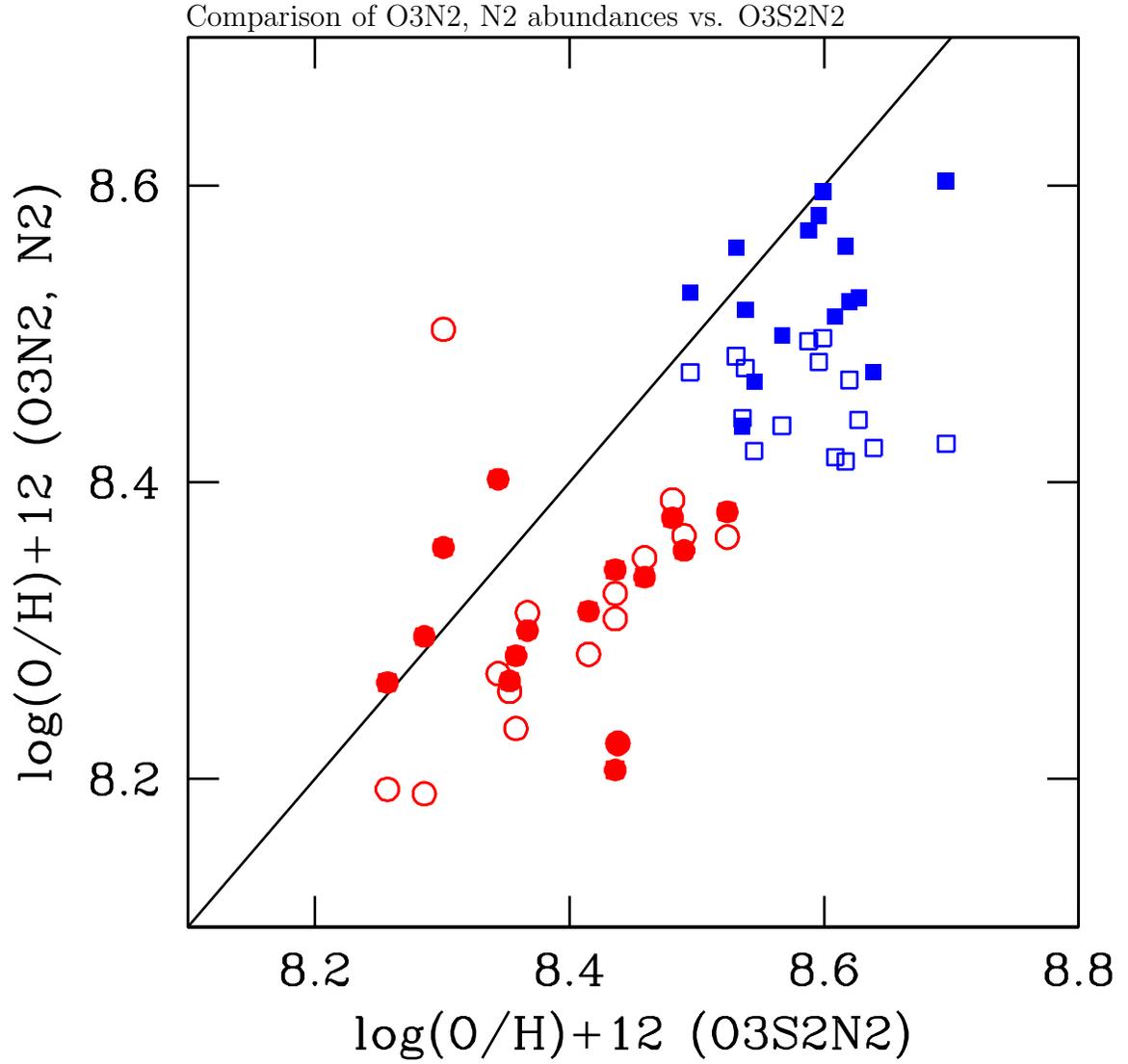}
\caption{Comparison between abundances from the O3N2 and N2 indexes (Marino et al. 2013) vs. those from OP3N2S2. Circles indicate NGC~7793, squares NGC~4945 H~II regions. Filled symbols: O3N2 vs. O3S2N2; open symbols: N2 vs. O3S2N2.  The solid line represents the 1:1 relation. }
\end{figure}

\clearpage

\begin{figure}
\centering
\title{NGC~7793 radial oxygen gradients}
\includegraphics[width=\hsize]{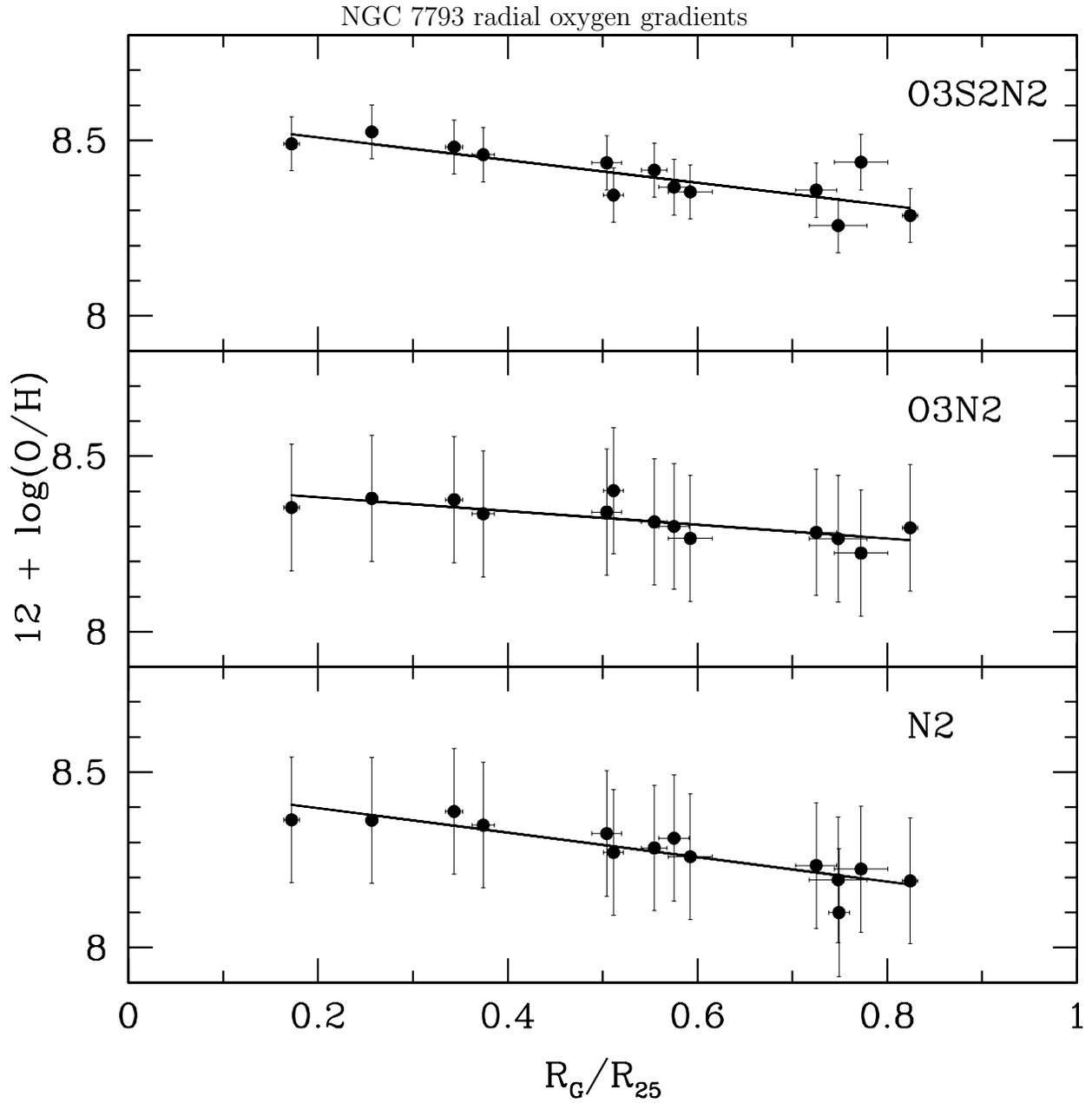}
\caption{Radial oxygen gradients with NGC~7793 H~II region abundances. Top panel: Oxygen abundances calculated with the O3S2N2 calibration; middle panel: O3N2 calibration;  bottom panel: N2 calibration.The solid lines correspond to the fits to the data points, as explained in the text.}
\end{figure}

\begin{figure}
\centering
\title{NGC~4945 radial oxygen gradients}
\includegraphics[width=\hsize]{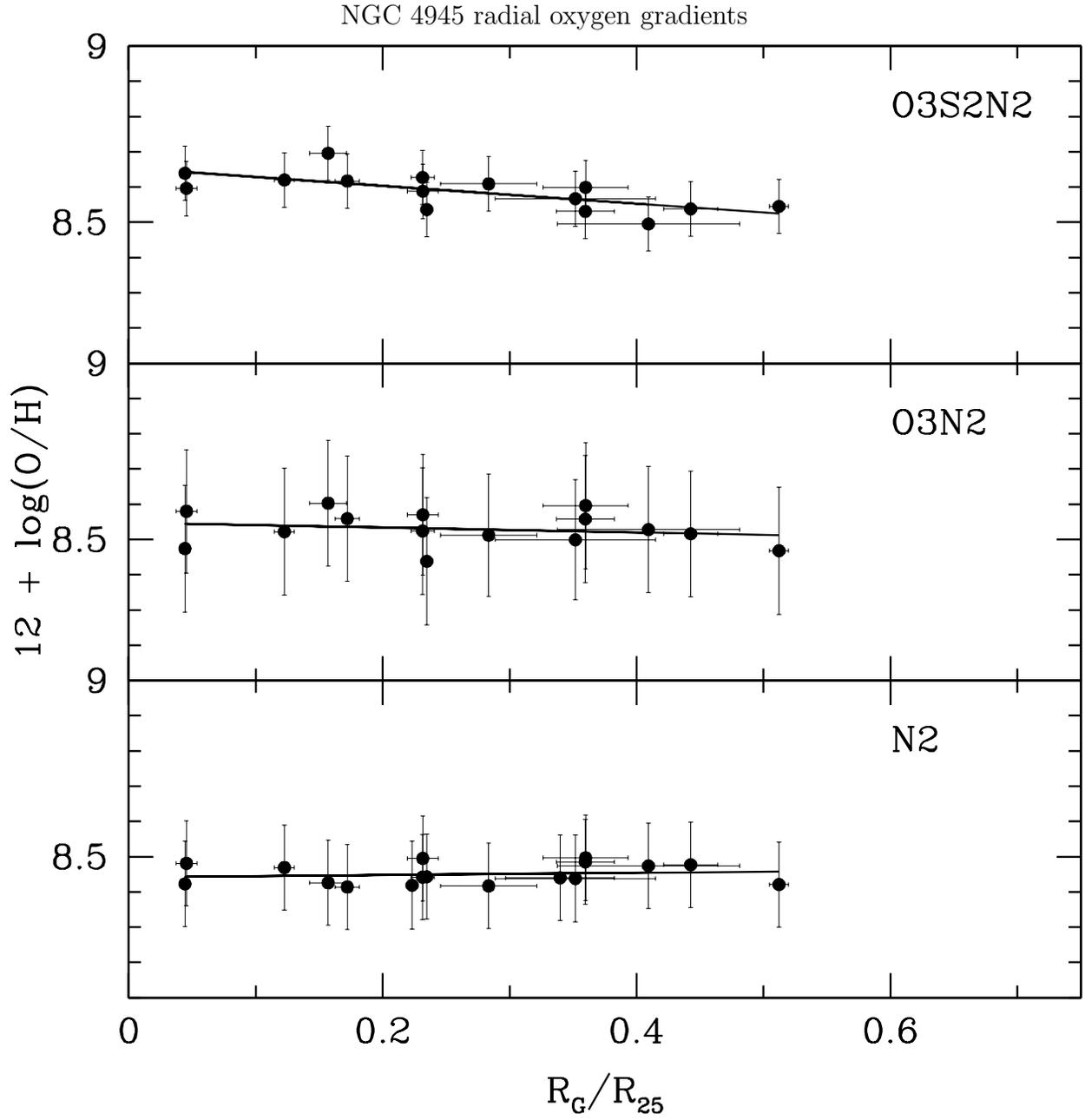}
\caption{Same as Fig. 6, but for NGC~4945 H~II regions.}
\end{figure}

\begin{figure}
\centering
\title{O3S2N2 and T$_{\rm e}$ abundances vs. O3N2, in M33 and M81}
\includegraphics[width=\hsize]{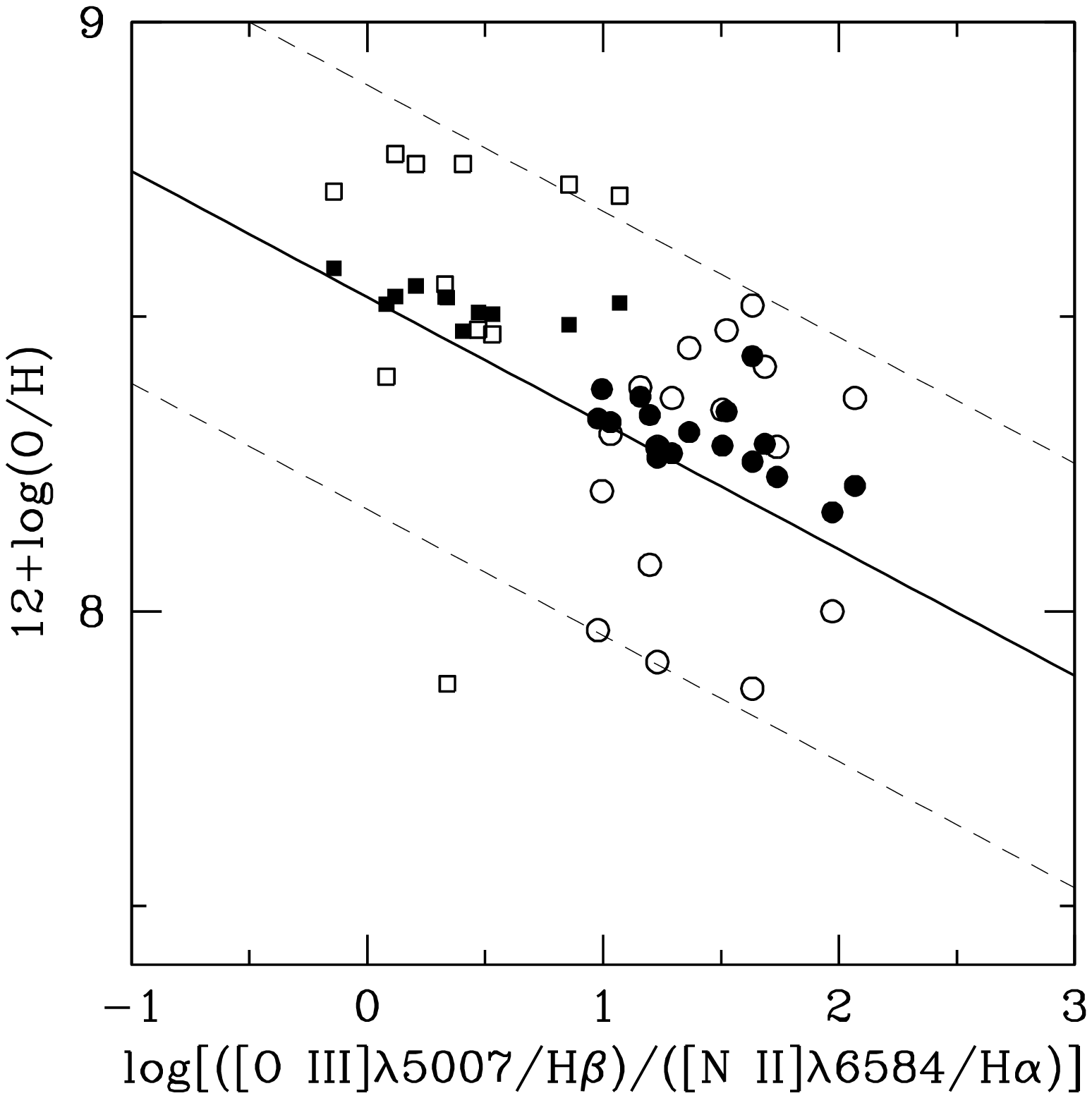}
\caption{Direct (T$_{\rm e}$, open symbols) and O3S2N2 (filled symbols) abundances against the O3N2 index. Circles: M33 H~II, squares: M81 H~II regions. The solid line is the Marino et al's calibration, the broken lines are the  $\pm$ 0.36 dex limits. }
\end{figure}

\clearpage

\begin{figure}
\centering
\title{O3S2N2 and T$_{\rm e}$ abundances vs. N2, in M33 and M81}
\includegraphics[width=\hsize]{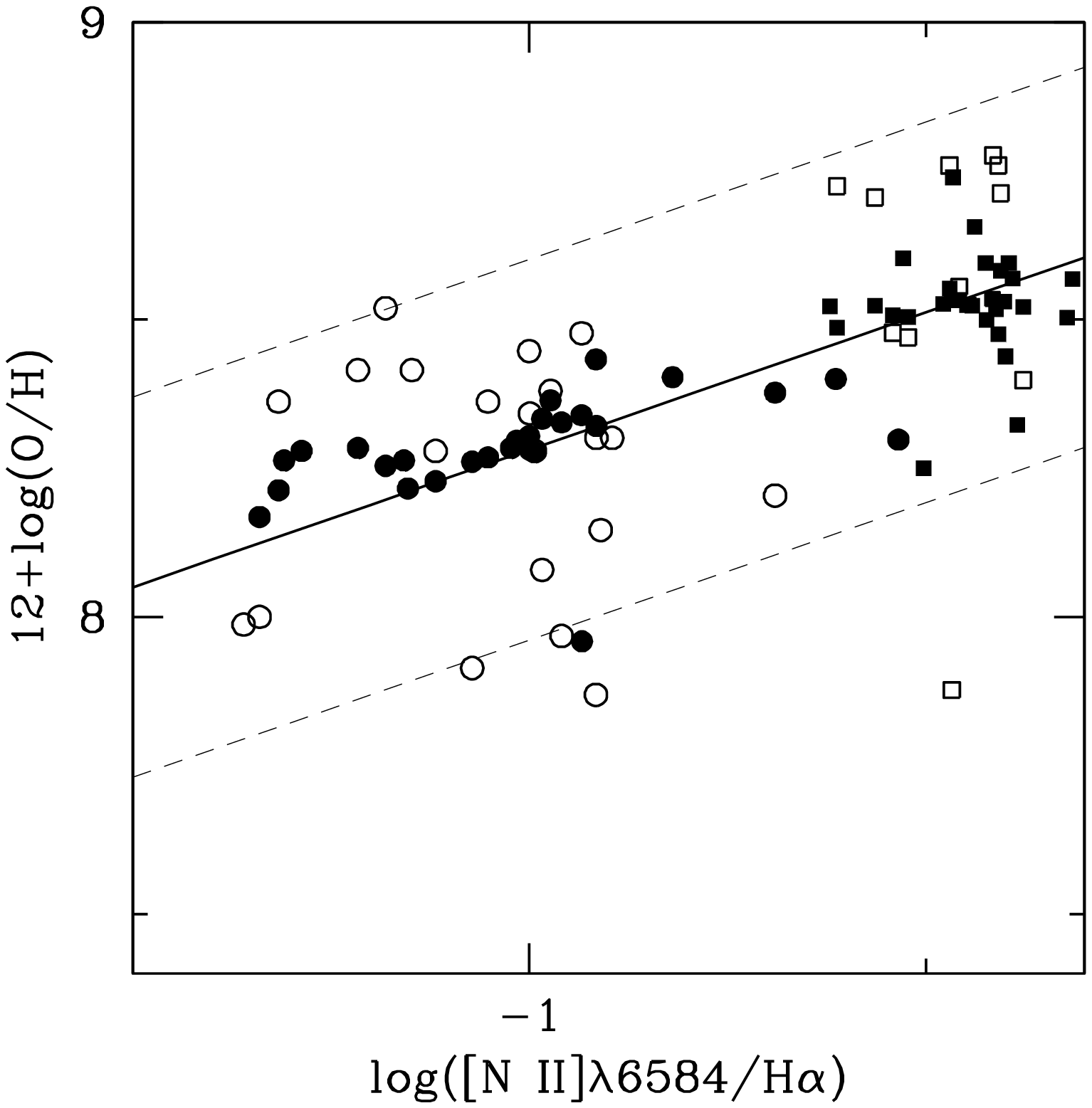}
\caption{Direct (T$_{\rm e}$, open symbols) and O3S2N2 (filled symbols) abundances against the N2 index. Circles: M33 H~II, squares: M81 H~II regions. The solid line is the Marino et al's  calibration, the broken lines are the  $\pm$ 0.32 dex limits.}
\end{figure}

\begin{figure}
\centering
\title{Radial oxygen gradients and twin-galaxy models}
\includegraphics[width=\hsize]{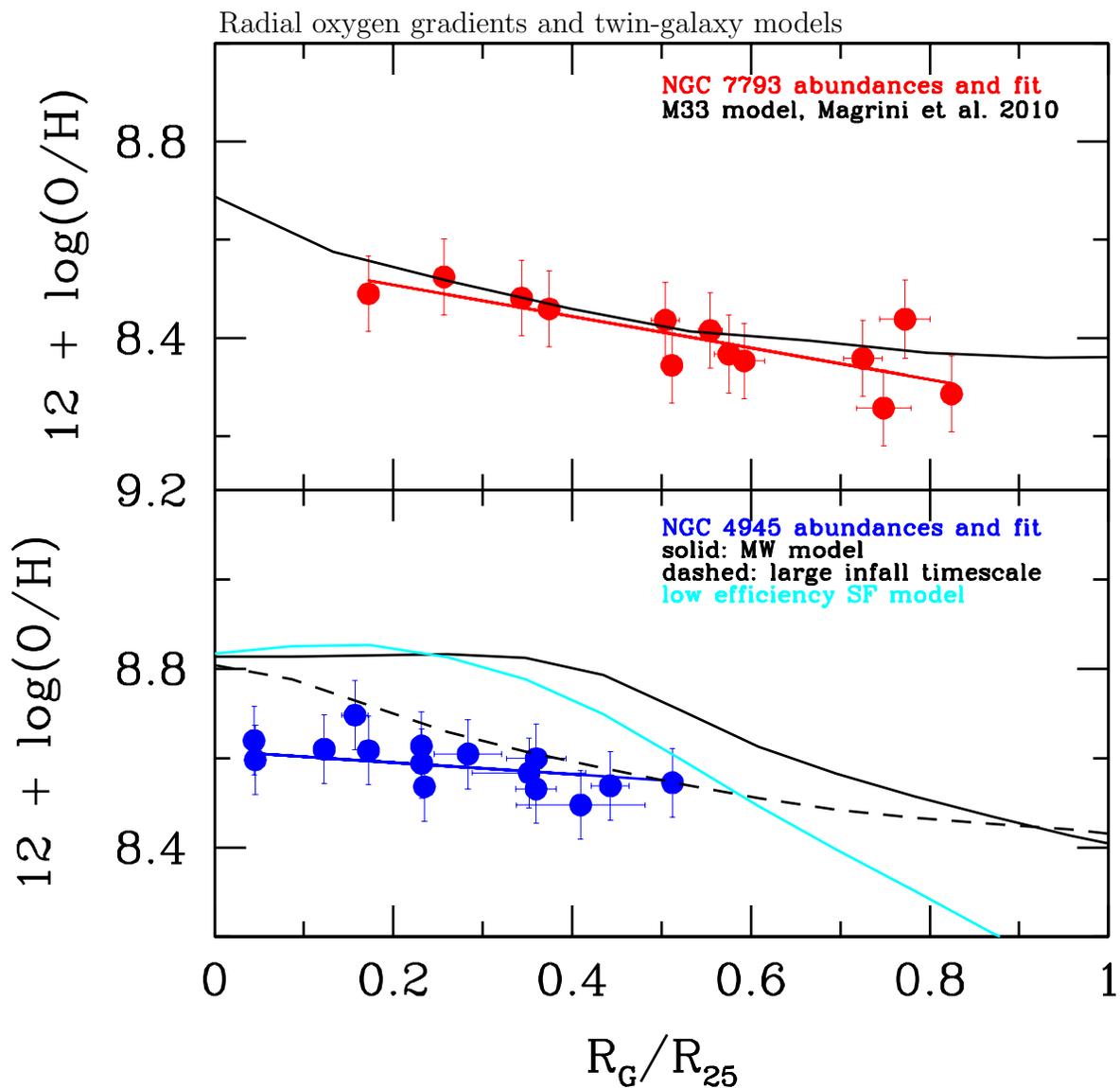}
\caption{Top panel: NGC~7793 gradient from O3S2N2 abundances, with their fit (blue dots and red solid line), and the M33 radial gradient (black line) from the model, as described in the text. Bottom panel: Same for NGC~4945, where the solid black line is the basic MW model, the light blue line is the MW model with low-efficiency star formation, and the dashed line is the MW model with larger infall time scale, as described in the text.}
\end{figure}

\clearpage









\clearpage

\begin{deluxetable}{lccccc}
\tablecaption{}

\tablehead{ \colhead{}&
\colhead{NGC~7793}&     \colhead{M33}&     \colhead{NGC~4945}& \colhead{MW}& \colhead{Ref.}}
\startdata

Type&	SA(s)d&	SA(s)cd&	SB(s)cd& 	SAB(rs)bc& \tablenotemark{a}\\

M$_{\rm V}$&			-20.01 $\pm$ 0.24&		 	-19.4&				-20.6&				-20.9&			\tablenotemark{b}\\

M [10$^{10}$ M$_{\odot}$]&	0.32&				0.49&				8 -- 15&				21&			\tablenotemark{c}\\

$\alpha_{\rm c, 2000}$&		23:57:49.8&			01:33:50.9&			13:05:27.5&			$\dots$&			\tablenotemark{b}\\

$\delta_{\rm c, 2000}$&		-32:35:27.7&			30:39:36.8&			-49:28:05.6&			$\dots$&			\tablenotemark{b}\\

R$_{\rm 25}$ [arcmin]\tablenotemark{d}&		5.24 $\pm$0.24& 30.8 $\pm$ 0.7&			11.72 $\pm$ 0.27&		$\dots$&			\tablenotemark{a}\\

Dist. [Mpc]&				3.91 $\pm$ 0.41&		840&				3.82 $\pm$ 0.31&		$\dots$&			\tablenotemark{e}\\

PA [deg]&					99 $\pm$ 1&			23 $\pm$ 5&			43 $\pm$ 1&			$\dots$&			\tablenotemark{f}\\

i [deg]&					53.7 $\pm$ 0.2&			54&					78 $\pm$ 3&			$\dots$&			\tablenotemark{f}\\
\enddata
\tablenotetext{a}{Hyperleda database, and references therein.}
\tablenotetext{b}{NED database, and references therein.}
\tablenotetext{c}{Leroy et al. 2008; Corbelli et al. 2014; Ables et al. 1987, and Dahlem et al. 1993; Kafle et al. 2012.}
\tablenotetext{d}{For the MW we used R$_{\rm 25}$=11.5 kpc from de Vaucouleurs \& Piece 1987.}
\tablenotetext{e}{Karachentsev et al. 2003; Freedman et al. 1991; Lin et al. 2011.}
\tablenotetext{f}{Corbelli \& Schneider 1997; Ables et al. 1987, and Dahlem et al. 1993.}

\end{deluxetable}

\clearpage

\begin{deluxetable}{llll}
\tablecaption{Observing log}
\tablehead{\colhead{Program}&	 \colhead{Target}& \colhead{Type}& \colhead{t$_{\rm exp}$}\\
&&&\colhead{[hr]}}
\startdata
GS2012BQ082& NGC~7793& Imaging\tablenotemark{a} & 1.14 \\
GS2012BQ046& NGC~4945, F1& Imaging\tablenotemark{a} & 1.13\\
GS2012BQ046& NGC~4945, F2& Imaging\tablenotemark{a} & 1.13\\
GS2013AQ046&  NGC~7793& MOS B600& 3.36\tablenotemark{b}  \\
GS2013AQ046& NGC~7793& MOS R400& 2.26  \\
GS2013AQ046& NGC~4945, F1& MOS B600& 2.26\\
GS2013AQ046& NGC~4945, F1& MOS R400& 2.23\\
GS2013AQ046& NGC~4945, F2& MOS B600& 2.29\\
GS2013AQ046& NGC~4945, F2& MOS R400& 3.52\tablenotemark{b}\\
\enddata
\tablenotetext{a}{Imaging includes [O~III] $\lambda$5007, H$\alpha$ and continuum filters.}
\tablenotetext{b}{Few low quality observations were repeated. Total time includes repeats.}
\end{deluxetable}

\clearpage

\begin{deluxetable}{lcccc}
\tablecaption{Positions and distances of H~II regions in NGC~7793}
\tablehead{\colhead{Region} & \colhead{$\alpha_{\rm 2000}$} & \colhead{$\delta_{\rm 2000}$}&  \colhead {R$_{\rm G}$/R$_{\rm 25}$}& \colhead{R$_{\rm G}$}\\
&&&&\colhead{[kpc]}}
\startdata
                  2&        23:58:00.51&        -32:33:20.09 & 0.87 $\pm$ 0.04 & 5.20 $\pm$ 0.22 \\ 
                  3&        23:57:58.64&        -32:34:33.27 & 0.50 $\pm$ 0.02 & 3.01 $\pm$ 0.09 \\ 
                  4\tablenotemark{a} &        23:58:01.84&        -32:34:20.97 & 0.66 $\pm$ 0.02 & 3.93 $\pm$ 0.11 \\ 
                  5\tablenotemark{b} &        23:57:58.32&        -32:34:09.05 & 0.59 $\pm$ 0.02 & 3.53 $\pm$ 0.13 \\ 
                  6&        23:58:01.73&        -32:33:51.02 & 0.77 $\pm$ 0.03 & 4.60 $\pm$ 0.16 \\ 
                  7\tablenotemark{a} &        23:57:52.05&        -32:34:05.74 & 0.46 $\pm$ 0.02 & 2.75 $\pm$ 0.14 \\ 
                  8&        23:58:02.96&        -32:34:13.66 & 0.72 $\pm$ 0.02 & 4.32 $\pm$ 0.12 \\ 
                  9&        23:57:59.91&        -32:33:43.89 & 0.75 $\pm$ 0.03 & 4.46 $\pm$ 0.18 \\ 
                 10\tablenotemark{a} &        23:57:49.65&        -32:33:59.58 & 0.47 $\pm$ 0.02 & 2.79 $\pm$ 0.14 \\ 
                 11&        23:58:06.81&        -32:34:57.25 & 0.75 $\pm$ 0.01 & 4.46 $\pm$ 0.05 \\ 
                 12&        23:58:00.54&        -32:34:52.47 & 0.51 $\pm$ 0.01 & 3.05 $\pm$ 0.06 \\ 
                 13\tablenotemark{b} &        23:58:00.19&        -32:34:45.73 & 0.52 $\pm$ 0.01 & 3.08 $\pm$ 0.07 \\ 
                 14\tablenotemark{c} &        23:58:12.72&        -32:35:50.38 & 0.93 $\pm$ 0.01 & 5.57 $\pm$ 0.01 \\ 
                 15&        23:57:56.08&        -32:35:39.77 & 0.26 $\pm$ 0.00 & 1.53 $\pm$ 0.00 \\ 
                 16\tablenotemark{d}  &        23:57:51.73&        -32:35:26.95 & 0.08 $\pm$ 0.00 & 0.48 $\pm$ 0.00 \\ 
                 17\tablenotemark{c}  &        23:57:58.93&        -32:35:19.10 & 0.39 $\pm$ 0.00 & 2.31 $\pm$ 0.02 \\ 
                 18&        23:58:08.79&        -32:36:46.10 & 0.82 $\pm$ 0.01 & 4.91 $\pm$ 0.02 \\ 
                 19&        23:57:58.72&        -32:37:03.34 & 0.57 $\pm$ 0.02 & 3.43 $\pm$ 0.09 \\ 
                 20&        23:57:59.07&        -32:36:56.26 & 0.55 $\pm$ 0.01 & 3.30 $\pm$ 0.08 \\ 
                 21&        23:57:55.11&        -32:36:32.10 & 0.37 $\pm$ 0.01 & 2.23 $\pm$ 0.07 \\ 
                 22&        23:57:50.55&        -32:36:00.49 & 0.17 $\pm$ 0.01 & 1.03 $\pm$ 0.05 \\ 
                 23\tablenotemark{e} &        23:57:57.06&        -32:35:55.05 & 0.31 $\pm$ 0.00 & 1.85 $\pm$ 0.01 \\ 
                 24&        23:57:55.39&        -32:36:23.44 & 0.34 $\pm$ 0.01 & 2.05 $\pm$ 0.05 \\ 
                 25\tablenotemark{f}  &        23:57:56.41&        -32:36:07.16 & 0.31 $\pm$ 0.00 & 1.87 $\pm$ 0.02 \\ 
                 26\tablenotemark{f} &        23:57:57.92&        -32:36:37.79 & 0.46 $\pm$ 0.01 & 2.72 $\pm$ 0.05 \\ 
                 27\tablenotemark{b} &        23:57:58.40&        -32:36:16.24 & 0.40 $\pm$ 0.00 & 2.39 $\pm$ 0.02 \\ 
                 28&        23:57:55.90&        -32:37:16.79 & 0.59 $\pm$ 0.02 & 3.53 $\pm$ 0.13 \\

\enddata
\tablenotetext{a}{Very poor spectrum, not used for abundance analysis.}
\tablenotetext{b} {Poor spectrum, low S/N, not included in gradients.}
\tablenotetext{c}{H~II region nature possible, but it could also be a PN. On BPT boundary. Not included in gradients.}
\tablenotetext{d}{H~II nature very probable. On BPT boundary. not included in gradients.}
\tablenotetext{e}{~II region nature possible, but it could also be a PN. BPT outlier. Not included in gradients.}
\tablenotetext{f}{H~II nature very probable. BPT outlier. Not included in gradients.}

\end{deluxetable}

\begin{deluxetable}{lcccc}
\tablecaption{Positions and distances of H~II regions in NGC~4945}
 \tablehead{\colhead{Region} & \colhead{$\alpha_{\rm 2000}$} & \colhead{$\delta_{\rm 2000}$}& \colhead{R$_{\rm G}$/R$_{\rm 25}$}& \colhead{R}\\
 &&&&\colhead{[kpc]}}
\startdata 
Field 1&&&&\\
&&&&\\

1&  13:05:43.31&  -49:24:40.00& 0.41 $\pm$ 0.02& 5.34 $\pm$ 0.31\\
 2\tablenotemark{a}&  13:05:47.16&  -49:24:08.30& 0.46 $\pm$ 0.02& 5.97 $\pm$ 0.23\\
 3\tablenotemark{a}&  13:05:50.08&  -49:23:12.24& 0.58 $\pm$ 0.03& 7.61 $\pm$ 0.44\\
 4&  13:05:48.94&  -49:23:50.64& 0.49 $\pm$ 0.02& 6.39 $\pm$ 0.22\\
 5&  13:05:51.75&  -49:23:38.17& 0.51 $\pm$ 0.01& 6.67 $\pm$ 0.09\\
 6&  13:05:45.57&  -49:24:19.49& 0.44 $\pm$ 0.02& 5.77 $\pm$ 0.28\\ 
 7\tablenotemark{a}&  13:05:45.79&  -49:24:50.76& 0.38 $\pm$ 0.00& 4.90 $\pm$ 0.03\\ 
 8&  13:05:47.37&  -49:25:31.45& 0.43 $\pm$ 0.03& 5.57 $\pm$ 0.44\\ 
 9&  13:05:44.85&  -49:25:42.38& 0.36 $\pm$ 0.02& 4.68 $\pm$ 0.30\\
10\tablenotemark{a}&  13:05:43.39&  -49:26:28.10& 0.41 $\pm$ 0.05& 5.29 $\pm$ 0.65\\
11&  13:05:36.77&  -49:26:08.22& 0.23 $\pm$ 0.01& 3.02 $\pm$ 0.16\\
12&  13:05:34.25&  -49:26:39.06& 0.17 $\pm$ 0.01& 2.24 $\pm$ 0.12\\ 
 &&&&\\
Field 2&&&&\\
 &&&&\\
 1&  13:05:15.92&  -49:29:50.46& 0.23 $\pm$ 0.01& 3.01 $\pm$ 0.12\\
 2&  13:05:17.14&  -49:30:02.91& 0.22 $\pm$ 0.004& 2.91 $\pm$ 0.06\\ 
 3&  13:05:20.38&  -49:30:18.50& 0.34 $\pm$ 0.04& 4.43 $\pm$ 0.56\\ 
 4&  13:05:27.88&  -49:28:11.78& 0.05 $\pm$ 0.01& 0.59 $\pm$ 0.11\\ 
 5&  13:05:27.07&  -49:29:39.96& 0.43 $\pm$ 0.07& 5.60 $\pm$ 0.97\\ 
 6&  13:05:28.00&  -49:29:26.86& 0.41 $\pm$ 0.07& 5.33 $\pm$ 0.93\\ 
 7\tablenotemark{a}&  13:05:31.86&  -49:28:25.79& 0.31 $\pm$ 0.06& 3.98 $\pm$ 0.72\\
 8&  13:05:23.44&  -49:28:40.83& 0.08 $\pm$ 0.004& 1.08 $\pm$ 0.06\\ 
 9\tablenotemark{a}&  13:05:28.82&  -49:29:07.35& 0.35 $\pm$ 0.06& 4.61 $\pm$ 0.82\\ 
10\tablenotemark{b}&  13:05:29.80&  -49:28:57.05& 0.35 $\pm$ 0.06& 4.58 $\pm$ 0.82\\ 
11\tablenotemark{a}&  13:05:27.95&  -49:29:16.33& 0.36 $\pm$ 0.06& 4.65 $\pm$ 0.81\\ 
12&  13:05:29.73&  -49:27:44.90& 0.04 $\pm$ 0.001& 0.58 $\pm$ 0.02\\ 
13&  13:05:38.38&  -49:27:24.01& 0.37 $\pm$ 0.06& 4.76 $\pm$ 0.75\\ 
14&  13:05:32.27&  -49:26:54.32& 0.16 $\pm$ 0.01& 2.05 $\pm$ 0.19\\ 
15&  13:05:32.08&  -49:27:05.11& 0.12 $\pm$ 0.01& 1.60 $\pm$ 0.10\\ 
16&  13:05:28.82&  -49:26:40.39& 0.35 $\pm$ 0.06& 4.56 $\pm$ 0.77\\ 
17\tablenotemark{a}&  13:05:42.66&  -49:26:29.42& 0.38 $\pm$ 0.04& 4.92 $\pm$ 0.58\\ 
18\tablenotemark{b}&  13:05:32.74&  -49:26:20.89& 0.28 $\pm$ 0.04& 3.69 $\pm$ 0.49\\ 
19&  13:05:39.31&  -49:26:10.36& 0.23 $\pm$ 0.004& 3.06 $\pm$ 0.05\\ 
20\tablenotemark{c}&  13:05:34.99&  -49:25:39.60& 0.39 $\pm$ 0.05& 5.05 $\pm$ 0.66\\ 
21&  13:05:38.42&  -49:25:21.81& 0.36 $\pm$ 0.03& 4.69 $\pm$ 0.43\\ 

  \enddata
\tablenotetext{a}{{Low S/N spectrum, not included in gradients.}}
\tablenotetext{b}{[S~II] emission lines have 2.5$<$S/N$<$3.}
\tablenotetext{c}{Probably not a H~II region. BPT outlier. Not included in gradients.}
\end{deluxetable}

\clearpage

\begin{deluxetable}{clrrr}
\tablecaption{NGC~7793 fluxes}
\tablehead{
\colhead{$\lambda$}&	\colhead{ID}&  \colhead{F}& \colhead{$\Delta$F}& \colhead{I}\\
\colhead{[\AA]}& & \colhead{[erg cm$^{-2}$ s$^{-1}$]}&  \colhead{[erg cm$^{-2}$ s$^{-1}$]}&  \colhead{[erg cm$^{-2}$ s$^{-1}$]}}
\startdata
Region 2, c=0. F$_{\rm H\beta}$=2.70e-15$\pm$5e-17 \\
H$\beta$    &     4861&     100.0 &       1.922 &         100.0 \\  
$\dots$& $\dots$& $\dots$& $\dots$& $\dots$\\
\enddata 
\tablecomments{This Table is published in its entirety in the electronic edition of the Astrophysical Journal. A portion is shown here for guidance regarding its form and content.}
\end{deluxetable}

\begin{deluxetable}{clrrr}
\tablecaption{NGC~4945 fluxes}
\tablehead{
\colhead{$\lambda$}&	\colhead{ID}&  \colhead{F}& \colhead{$\Delta$F}& \colhead{I}\\
\colhead{[\AA]}& & \colhead{[erg cm$^{-2}$ s$^{-1}$]}&  \colhead{[erg cm$^{-2}$ s$^{-1}$]}& \colhead{[erg cm$^{-2}$ s$^{-1}$]}}
\startdata
Field 1\\
Region 1, c=0.65, F$_{\rm H\beta}$=4.446e-15$\pm$1e-17\\
H$\beta$&     4861   &  100.0    &    0.382    &      100.0   \\
$\dots$&$\dots$& $\dots$&$\dots$&$\dots$\\
\enddata 
\tablecomments{This Table is published in its entirety in the electronic edition of the Astrophysical Journal. A portion is shown here for guidance regarding its form and content.}
\end{deluxetable}
 
\clearpage

\clearpage

\begin{deluxetable}{lccc}
\tablecaption{Strong-line oxygen abundances of NGC~7793 regions}

\tablehead{ \colhead{Region}&
\colhead{12+log(O/H)}  &     \colhead{12+log(O/H)}&     \colhead{12+log(O/H)}\\
\colhead{}& \colhead{O3S2N2}& \colhead{O3N2}& \colhead{N2}}

\startdata
 3 & 8.436$\pm$0.077 & 8.341$\pm$0.180 & 8.325$\pm$0.179 \\
 4 & $\dots$ & $\dots$ & 8.325$\pm$0.275 \\
 5 & 8.300$\pm$0.079 & 8.334$\pm$0.179 & 8.298$\pm$0.180 \\
 6 & 8.438$\pm$0.080 & 8.224$\pm$0.179 & 8.224$\pm$0.180 \\
 7 & $\dots$ & 8.407$\pm$0.077 & $\dots$ \\
 8 & 8.358$\pm$0.078 & 8.283$\pm$0.180 & 8.234$\pm$0.179 \\
 9 & 8.257$\pm$0.077 & 8.265$\pm$0.180 & 8.193$\pm$0.179 \\
10 & $\dots$ & 8.261$\pm$0.083 & 8.518$\pm$0.236 \\
11 & $\dots$ & $\dots$ & 8.100$\pm$0.182 \\
12 & 8.344$\pm$0.077 & 8.402$\pm$0.180 & 8.271$\pm$0.179 \\
13 & 8.370$\pm$0.077 & 8.203$\pm$0.180 & 8.208$\pm$0.179 \\
14 & $\dots$ & 8.285$\pm$0.176 & 8.404$\pm$0.180 \\
15 & 8.524$\pm$0.077 & 8.380$\pm$0.180 & 8.363$\pm$0.179 \\
16 & 8.301$\pm$0.077 & 8.356$\pm$0.179 & 8.503$\pm$0.180 \\
17 & $\dots$ & $\dots$ & 8.232$\pm$0.194 \\
18 & 8.286$\pm$0.077 & 8.296$\pm$0.180 & 8.190$\pm$0.179 \\
19 & 8.367$\pm$0.079 & 8.300$\pm$0.179 & 8.312$\pm$0.180 \\
20 & 8.415$\pm$0.077 & 8.313$\pm$0.180 & 8.284$\pm$0.179 \\
21 & 8.459$\pm$0.077 & 8.336$\pm$0.180 & 8.349$\pm$0.179 \\
22 & 8.490$\pm$0.077 & 8.354$\pm$0.180 & 8.364$\pm$0.179 \\
23 & $\dots$ & 8.310$\pm$0.180 & 8.486$\pm$0.179 \\
24 & 8.481$\pm$0.077 & 8.376$\pm$0.180 & 8.388$\pm$0.179 \\
25 & 8.436$\pm$0.077 & 8.206$\pm$0.180 & 8.308$\pm$0.179 \\
26 & $\dots$ & 8.240$\pm$0.169 & 8.369$\pm$0.188 \\
27 & 8.444$\pm$0.077 & 8.434$\pm$0.180 & 8.372$\pm$0.179 \\
28 & 8.353$\pm$0.077 & 8.266$\pm$0.180 & 8.259$\pm$0.179 \\

\enddata
\end{deluxetable}

\begin{deluxetable}{lccc}
\tablecaption{Strong-line oxygen abundances of NGC~4945 regions}

\tablehead{ \colhead{Region}&
\colhead{12+log(O/H)}  &     \colhead{12+log(O/H)}&     \colhead{12+log(O/H)}\\
\colhead{}& \colhead{O3S2N2}& \colhead{O3N2}& \colhead{N2}}

\startdata
Field 1\\
 2 & 8.334 $\pm$0.077 & 8.670 $\pm$0.180 &  $\dots$ \\
 5 & 8.545 $\pm$0.077 & 8.468 $\pm$0.180 & 8.421 $\pm$0.120 \\
 6 & 8.538 $\pm$0.077 & 8.516 $\pm$0.178 & 8.477 $\pm$0.121 \\
 7 &  $\dots$ &  $\dots$ & 8.401 $\pm$0.123 \\
 9 & 8.531 $\pm$0.077 & 8.558 $\pm$0.180 & 8.485 $\pm$0.120 \\
 10 & 8.624 $\pm$0.077 & 8.511 $\pm$0.179 & 8.417 $\pm$0.121 \\
 11 & 8.588 $\pm$0.077 & 8.570 $\pm$0.171 & 8.495 $\pm$0.121 \\
 12 & 8.617 $\pm$0.077 & 8.559 $\pm$0.178 & 8.414 $\pm$0.121 \\

Field2\\
    1 & 8.627 $\pm$0.077 & 8.524 $\pm$0.180 & 8.442 $\pm$0.120 \\
    2 &  $\dots$ &  $\dots$ & 8.419 $\pm$0.125 \\
    3 &  $\dots$ &  $\dots$ & 8.440 $\pm$0.121 \\
    4 & 8.596 $\pm$0.077 & 8.580 $\pm$0.175 & 8.481 $\pm$0.121 \\
    6 & 8.495 $\pm$0.077 & 8.528 $\pm$0.179 & 8.474 $\pm$0.121 \\
    7 & 8.439 $\pm$0.082 & 8.365 $\pm$0.139 & 8.429 $\pm$0.161 \\
    9 & 8.578 $\pm$0.078 & 8.632 $\pm$0.144 & 8.402 $\pm$0.123 \\
    10 & 8.567 $\pm$0.078 & 8.499 $\pm$0.170 & 8.438 $\pm$0.123 \\
    11 & 8.626 $\pm$0.078 & 8.601 $\pm$0.145 & 8.485 $\pm$0.123 \\
    12 & 8.639 $\pm$0.077 & 8.474 $\pm$0.180 & 8.423 $\pm$0.121 \\
    14 & 8.696 $\pm$0.077 & 8.603 $\pm$0.179 & 8.426 $\pm$0.121 \\
    15 & 8.620 $\pm$0.077 & 8.522 $\pm$0.180 & 8.469 $\pm$0.120 \\
    17 & 8.344 $\pm$0.135 & 8.473 $\pm$0.153 & 8.237 $\pm$0.142 \\
    18 & 8.609 $\pm$0.077 & 8.512 $\pm$0.174 & 8.417 $\pm$0.121 \\
    19 & 8.536 $\pm$0.077 & 8.438 $\pm$0.180 & 8.443 $\pm$0.120 \\
    20 & 8.249 $\pm$0.077 & 8.425 $\pm$0.178 &  $\dots$ \\
    21 & 8.599 $\pm$0.077 & 8.596 $\pm$0.179 & 8.497 $\pm$0.121 \\

\enddata
\end{deluxetable}

\begin{deluxetable}{lrrrrrr}
\tablecaption{Radial oxygen gradients in NGC~7793 and NGC~4945 from strong-line abundances of H~II regions}

\tablehead{ \colhead{Method}&
\colhead{N}  &     
\colhead{Range}&     

\colhead{Slope}&
\colhead{Slope}&
\colhead{Intercept}&
\colhead{$<$O/H$>$}\\

\colhead{}&
\colhead{}& 
\colhead{[R$_{\rm25}$]}&
\colhead{[dex R$_{\rm 25}^{-1}$]}&
\colhead{[dex kpc$^{-1}$]}&
\colhead{[dex]}&
\colhead{[10$^{-4}$]}\\
}

\startdata
NGC~7793&&&&&\\
\hline

O3S2N2&	13&	0.17--0.82&	-0.321 $\pm$ 0.112&	 	-0.054 $\pm$ 0.019& 	8.572 $\pm$ 0.063& 		2.55 $\pm$ 0.46\\
O3N2&	13&	0.17--0.82&	-0.195 $\pm$ 0.276&		-0.033 $\pm$ 0.046&		8.422 $\pm$ 0.145& 		2.09 $\pm$ 0.25\\
N2&		14&	0.17--0.82&	-0.350 $\pm$ 0.263&		-0.059 $\pm$ 0.044&		8.468 $\pm$ 0.142& 		1.92 $\pm$ 0.35\\

&&&&&\\
NGC~4945&&&&&\\
\hline

O3S2N2&	15&	0.04--0.51&	-0.253 $\pm$ 0.149&		-0.019 $\pm$ 0.011&		8.654 $\pm$ 0.041&		3.89 $\pm$ 0.47\\
O3N2&	15&	0.04--0.51&	-0.070 $\pm$ 0.275&		-0.005 $\pm$ 0.021&		8.548 $\pm$ 0.100&		3.41 $\pm$ 0.37\\
N2&		19&	0.04--0.51&	0.031 $\pm$0.179&		0.002 $\pm$ 0.014&		8.442 $\pm$0.067&		2.83 $\pm$ 0.19\\

\enddata
\end{deluxetable}

\clearpage


\begin{thebibliography}{}

\bibitem[Ables et al.(1987)]{1987MNRAS.226..157A} Ables, J.~G., Forster, J.~R., Manchester, R.~N., et al.\ 1987, \mnras, 226, 157 
\bibitem[Athanassoula(1992)]{1992MNRAS.259..328A} Athanassoula, E.\ 1992, \mnras, 259, 328 
\bibitem[Baldwin et al.(1981)]{1981PASP...93....5B} Baldwin, J.~A., Phillips, M.~M., \& Terlevich, R.\ 1981, \pasp, 93, 5 
\bibitem[Berg et al.(2013)]{2013ApJ...775..128B} Berg, D.~A., Skillman, E.~D., Garnett, D.~R., et al.\ 2013, \apj, 775, 128 
\bibitem[Bibby \& Crowther(2010)]{2010MNRAS.405.2737B} Bibby, J.~L., \& Crowther, P.~A.\ 2010, \mnras, 405, 2737 
\bibitem[Braatz et al.(1997)]{1997ApJS..110..321B} Braatz, J.~A., Wilson,  A.~S., \& Henkel, C.\ 1997, \apjs, 110, 321    
\bibitem[Bresolin et al.(2009)]{2009ApJ...700..309B} Bresolin, F., Gieren, W., Kudritzki, R.-P., et al.\ 2009, \apj, 700, 309 
\bibitem[Bresolin(2011)]{2011ApJ...730..129B} Bresolin, F.\ 2011, \apj, 730, 129 
\bibitem[Casasola et al.(2004)]{2004A&A...422..941C} Casasola, V., Bettoni, D., \& Galletta, G.\ 2004, \aap, 422, 941 
\bibitem[Casasola et  al.(2008)]{2008A&A...490...61C} Casasola, V., Combes, F., Garc{\'{\i}}a-Burillo, S., et al.\ 2008, \aap, 490, 61    
\bibitem[Casasola et al.(2011)]{2011A&A...527A..92C} Casasola, V., Hunt, L.~K., Combes, F., Garc{\'{\i}}a-Burillo, S., \& Neri, R.\ 2011, \aap, 527, AA92 
\bibitem[Charlot \& Longhetti(2001)]{2001MNRAS.323..887C} Charlot, S., \& Longhetti, M.\ 2001, \mnras, 323, 887 
\bibitem[Chou et al.(2007)]{2007ApJ...670..116C} Chou, R.~C.~Y., Peck,  A.~B., Lim, J., et al.\ 2007, \apj, 670, 116    
\bibitem[Combes et  al.(2013)]{2013A&A...558A.124C} Combes, F., Garc{\'{\i}}a-Burillo, S., Casasola, V., et al.\ 2013, \aap, 558, A124    
\bibitem[Corbelli et al.(2014)]{2014A&A...572A..23C} Corbelli, E., Thilker, D., Zibetti, S., Giovanardi, C., \& Salucci, P.\ 2014, \aap, 572, A23 
\bibitem[C{\^o}t{\'e} et al.(2009)]{2009AJ....138.1037C} C{\^o}t{\'e}, S., Draginda, A., Skillman, E.~D., \& Miller, B.~W.\ 2009, \aj, 138, 1037 
\bibitem[Curran et  al.(2001)]{2001A&A...367..457C} Curran, S.~J., Johansson, L.~E.~B., Bergman, P., Heikkil{\"a}, A., \& Aalto, S.\ 2001, \aap, 367, 457    
\bibitem[Dahlem et al.(1993)]{1993A&A...270...29D} Dahlem, M., Golla, G., Whiteoak, J.~B., et al.\ 1993, \aap, 270, 29 
\bibitem[Dalcanton(2007)]{2007ApJ...658..941D} Dalcanton, J.~J.\ 2007, \apj, 658, 941 
\bibitem[Deharveng et al.(2000)]{2000MNRAS.311..329D} Deharveng, L., Pe{\~n}a, M., Caplan, J., \& Costero, R.\ 2000, \mnras, 311, 329 
\bibitem[Denicol{\'o} et al.(2002)]{2002MNRAS.330...69D} Denicol{\'o}, G., Terlevich, R., \& Terlevich, E.\ 2002, \mnras, 330, 69 
\bibitem[Done et al.(1996)]{1996ApJ...463L..63D} Done, C., Madejski, G.~M.,  \& Smith, D.~A.\ 1996, \apjl, 463, L63   
\bibitem[Ferrini et al.(1992)]{1992ApJ...387..138F} Ferrini, F., Matteucci, F., Pardi, C., \& Penco, U.\ 1992, \apj, 387, 138 
\bibitem[Friedli et al.(1994)]{1994ApJ...430L.105F} Friedli, D., Benz, W., \& Kennicutt, R.\ 1994, \apjl, 430, L105 
\bibitem[Freedman et al.(1991)]{1991ApJ...372..455F} Freedman, W.~L., Wilson, C.~D., \& Madore, B.~F.\ 1991, \apj, 372, 455 
\bibitem[Gibson et al.(2013)]{2013A&A...554A..47G} Gibson, B.~K., Pilkington, K., Brook, C.~B., Stinson, G.~S., \& Bailin, J.\ 2013, \aap, 554, A47 
\bibitem[Guo\& Mathews(2012)]{2012ApJ...756..181G} Guo, F., \& Mathews, W.~G.\ 2012, \apj, 756, 181 
\bibitem[Heckman et al.(1990)]{1990ApJS...74..833H} Heckman, T.~M., Armus, L., \& Miley, G.~K.\ 1990, \apjs, 74, 833 
\bibitem[Henkel et  al.(1994)]{1994A&A...284...17H} Henkel, C., Whiteoak, J.~B., \& Mauersberger, R.\ 1994, \aap, 284, 17    
\bibitem[Iwasawa et al.(1993)]{1993ApJ...409..155I} Iwasawa, K., Koyama,  K., Awaki, H., et al.\ 1993, \apj, 409, 155   
\bibitem[Kafle et al.(2012)]{2012ApJ...761...98K} Kafle, P.~R., Sharma, S., Lewis, G.~F., \& Bland-Hawthorn, J.\ 2012, \apj, 761, 98 
\bibitem[Karachentsev et al.(2003)]{2003A&A...404...93K} Karachentsev, I.~D., Grebel, E.~K., Sharina, M.~E., et al.\ 2003, \aap, 404, 93 
\bibitem[Kewley \& Dopita(2002)]{2002ApJS..142...35K} Kewley, L.~J., \& Dopita, M.~A.\ 2002, \apjs, 142, 35 
\bibitem[Kniazev et al.(2008)]{2008MNRAS.384.1045K} Kniazev, A.~Y., Pustilnik, S.~A., \& Zucker, D.~B.\ 2008, \mnras, 384, 1045 
\bibitem[Leroy et al.(2008)]{2008AJ....136.2782L} Leroy, A.~K., Walter, F., Brinks, E., et al.\ 2008, \aj, 136, 2782 
\bibitem[Lin et al.(2011)]{2011ApJ...731...15L} Lin, L.-H., Taam, R.~E., Yen, D.~C.~C., Muller, S., \& Lim, J.\ 2011, \apj, 731, 15 
\bibitem[Madejski et al.(2000)]{2000ApJ...535L..87M} Madejski, G.,  {\.Z}ycki, P., Done, C., et al.\ 2000, \apjl, 535, L87    
\bibitem[Magrini et al.(2007)]{2007A&A...470..843M} Magrini, L., Corbelli, E., \& Galli, D.\ 2007, \aap, 470, 843 
\bibitem[Magrini et al.(2009a)]{2009ApJ...696..729M} Magrini, L., Stanghellini, L., \& Villaver, E.\ 2009, \apj, 696, 729 (2009a)
\bibitem[Magrini et al.(2009b)]{2009A&A...494...95M} Magrini, L., Sestito, P., Randich, S., \& Galli, D.\ 2009, \aap, 494, 95 (2009b)
\bibitem[Magrini et al.(2010)]{2010A&A...512A..63M} Magrini, L., Stanghellini, L., Corbelli, E., Galli, D., \& Villaver, E.\ 2010, \aap, 512, A63 
\bibitem[Marconi et  al.(2000)]{2000A&A...357...24M} Marconi, A., Oliva, E., van der Werf, P.~P., et al.\ 2000, \aap, 357, 24    
\bibitem[Marino et 
al.(2013)]{2013A&A...559A.114M} Marino, R.~A., Rosales-Ortega, F.~F., S{\'a}nchez, S.~F., et al.\ 2013, \aap, 559, A114 
\bibitem[Moustakas et al.(2010)]{2010ApJS..190..233M} Moustakas, J., Kennicutt, R.~C., Jr., Tremonti, C.~A., et al.\ 2010, \apjs, 190, 233 
\bibitem[]{}Narayanan, D. et al. 2008, ApJS, 176, 331
\bibitem[Onodera et al.(2004)]{2004PASJ...56..439O} Onodera, S., Koda, J., Sofue, Y., \& Kohno, K.\ 2004, \pasj, 56, 439 
\bibitem[Osterbrock \& Ferland(2006)]{2006agna.book.....O} Osterbrock, D.~E., \& Ferland, G.~J.\ 2006, Astrophysics of gaseous nebulae and active galactic nuclei, 2nd.~ed.~by D.E.~Osterbrock and G.J.~Ferland.~Sausalito, CA: University Science Books, 2006
\bibitem[Pe{\~n}a(2011)]{2011RMxAC..39...91P} Pe{\~n}a, M.\ 2011, Revista Mexicana de Astronomia y Astrofisica Conference Series, 39, 91 
\bibitem[Perez et al.(2011)]{2011MNRAS.417..580P} Perez, J., Michel-Dansac, L., \& Tissera, P.~B.\ 2011, \mnras, 417, 580 
\bibitem[Peterson(1980)]{1980PASP...92..397P} Peterson, C.~J.\ 1980, \pasp, 92, 397 
\bibitem[Pettini \& Pagel(2004)]{2004MNRAS.348L..59P} Pettini, M., \& Pagel, B.~E.~J.\ 2004, \mnras, 348, L59 
\bibitem[Pilkington et al.(2012)]{2012A&A...540A..56P} Pilkington, K., Few, C.~G., Gibson, B.~K., et al.\ 2012, \aap, 540, A56 
\bibitem[Pilyugin et al.(2014)]{2014AJ....147..131P} Pilyugin, L.~S., Grebel, E.~K., \& Kniazev, A.~Y.\ 2014, \aj, 147, 131 
\bibitem[Pilyugin \& Mattsson(2011)]{2011MNRAS.412.1145P} Pilyugin, L.~S., \& Mattsson, L.\ 2011, \mnras, 412, 1145 
\bibitem[Portinari \& Chiosi(2000)]{2000A&A...355..929P} Portinari, L., \& Chiosi, C.\ 2000, \aap, 355, 929 
\bibitem[Press et al. 1988]{Press}Press, W. et al., ``Numerical Recipes'' , 1988
\bibitem[Radburn-Smith et al.(2011)]{2011ApJS..195...18R} Radburn-Smith, D.~J., de Jong, R.~S., Seth, A.~C., et al.\ 2011, \apjs, 195, 18 
\bibitem[Rahimi et al.(2011)]{2011MNRAS.415.1469R} Rahimi, A., Kawata, D., Allende Prieto, C., et al.\ 2011, \mnras, 415, 1469 
\bibitem[S{\'a}nchez et al.(2014)]{2014A&A...563A..49S} S{\'a}nchez, S.~F., Rosales-Ortega, F.~F., Iglesias-P{\'a}ramo, J., et al.\ 2014, \aap, 563, AA49 
\bibitem[Schurch et al.(2002)]{2002MNRAS.335..241S} Schurch, N.~J.,  Roberts, T.~P., \& Warwick, R.~S.\ 2002, \mnras, 335, 241    
\bibitem[Stanghellini et al.(2014)]{2014A&A...567A..88S} Stanghellini, L., Magrini, L., Casasola, V., \& Villaver, E.\ 2014, \aap, 567, AA88 
\bibitem[Toomre \& Toomre(1972)]{1972ApJ...178..623T} Toomre, A., \& Toomre, J.\ 1972, \apj, 178, 623 
\bibitem[Van Dyk et al.(2012)]{2012AJ....143...19V} Van Dyk, S.~D., Davidge, T.~J., Elias-Rosa, N., et al.\ 2012, \aj, 143, 19 
\bibitem[Vila-Costas \& Edmunds(1992)]{1992MNRAS.259..121V} Vila-Costas, M.~B., \& Edmunds, M.~G.\ 1992, \mnras, 259, 121 
\bibitem[Wang et  al.(2004)]{2004A&A...422..883W} Wang, M., Henkel, C., Chin, Y.-N., et al.\ 2004, \aap, 422, 883    
\bibitem[Webster \& Smith(1983)]{1983MNRAS.204..743W} Webster, B.~L., \& Smith, M.~G.\ 1983, \mnras, 204, 743 
\bibitem[Weinmann et al.(2006)]{2006MNRAS.372.1161W} Weinmann, S.~M., van 
den Bosch, F.~C., Yang, X., et al.\ 2006, \mnras, 372, 1161 
\bibitem[Whiteoak \& Wilson(1990)]{1990MNRAS.245..665W} Whiteoak, J.~B., \& Wilson, W.~E.\ 1990, \mnras, 245, 665 




\end{thebibliography}
\end{document}